\documentclass[pra,onecolumn,floatfix,a4paper,superscriptaddress]{revtex4}
\usepackage{bm,color,graphicx,amsmath,txfonts}

\usepackage[colorlinks, citecolor=blue,linkcolor=blue]{hyperref}

\begin{document}

	\title{Nonreciprocal transparency windows, Fano resonance, and slow/fast light in a membrane-in-the-middle magnomechanical system induced by the Barnett effect}

	\author{M. Amghar}
	\affiliation{LPTHE-Department of Physics, Faculty of sciences, Ibnou Zohr University, Agadir, Morocco}

	\author{M. Amazioug}\email{m.amazioug@uiz.ac.ma}
	\affiliation{LPTHE-Department of Physics, Faculty of sciences, Ibnou Zohr University, Agadir, Morocco}

	\begin{abstract}
		
Nonreciprocal phenomena are currently a major focus of research within the fields of classical and quantum technology. In this work, we theoretically investigate the interplay among multiple magnomechanically induced transparency (MMIT) windows, Fano resonances, slow/fast light, and nonreciprocal absorption and group delay in a hybrid cavity magnomechanical system. This system is composed of two yttrium iron garnet (YIG) spheres and a membrane positioned at the center of the cavity. By analyzing the absorption spectrum of a weak probe field in the presence of a strong control field, we demonstrate the emergence of five transparency windows resulting from combined photon-phonon, photon-magnon, and phonon-magnon interactions. The photon-phonon coupling associated with the membrane plays a crucial role in enhancing and tailoring these transparency features. We further examine the impact of the Barnett effect on the absorption and dispersion characteristics, showing that it enables the controllable manipulation of transparency windows and the generation of tunable Fano resonance profiles. The influence of cavity decay and magnon dissipation rates on the spectral response is also analyzed. In addition, we demonstrate that the group delay of the transmitted probe field can be effectively tuned via the photon-phonon coupling strength and the Barnett effect, allowing for a controllable transition between slow and fast light regimes. Finally, nonreciprocal absorption and group delay are achieved through appropriate adjustment of the coupling parameters. These findings highlight the potential of the proposed hybrid system for applications in optical signal processing and quantum information technologies.

	\end{abstract}
	
	\date{\today}
	
	\maketitle
	\textbf{Keywords}: Nonreciprocal, transparency windows, Fano resonance, slow and fast light.
	%%%%%%%%%%%%%%%%%%%%%%%%%%%%%%%%%%%%%%%%%%%%%%%%
	\section{INTRODUCTION}
	Cavity magnomechanics \cite{0} has recently emerged as a rapidly advancing research field that focuses on the coherent interaction between electromagnetic fields and collective magnetic or mechanical excitations. These hybrid platforms offer promising opportunities for the development of next generation technologies, including quantum information processing, ultra-sensitive detection, and coherent signal transduction. Within this framework, microwave cavity systems provide a versatile and controllable environment for exploring photon–phonon–magnon interactions. Among the available material platforms, yttrium iron garnet (YIG) spheres have attracted remarkable interest due to their exceptionally low magnetic damping, strong magneto-optical coupling, and long coherence times, making them ideal candidates for realizing high performance cavity magnomechanical devices \cite{1,2,3}.\\
	
	A cavity magnomechanical system based on yttrium iron garnet (YIG) \cite{2} constitutes an advanced hybrid platform that integrates magnetic and mechanical degrees of freedom within a single resonant structure. In these systems, YIG spheres are commonly positioned inside a microwave cavity, where they act as the magnetic medium coupling simultaneously to the microwave field and mechanical vibrations. This architecture enables the investigation of rich physical phenomena arising from the interaction between YIG spin dynamics and mechanical motion, thereby establishing an effective framework for exploring magnomechanics. A broad range of effects have been reported in such platforms, including phase control of microwave transmission \cite{4}, generation of squeezed states \cite{5}, optical bistability \cite{6}, exceptional points \cite{7}, dynamical backaction \cite{8}, coherent coupling \cite{9}, nonreciprocal quantum phase transitions \cite{10}, magnetic sphere–microwave interactions \cite{11}, hybrid-system entanglement \cite{12,13}, cavity-mediated phenomena in microwave systems \cite{13,14,15,16}, nonreciprocal entanglement \cite{17,172}, Nonreciprocal quantum synchronization \cite{173}, and multiple magnomechanically induced transparency \cite{18,19,20}. These studies highlight the strong potential of cavity magnomechanical systems for applications in quantum information processing and precision sensing. Additional related works can be found in Refs. \cite{21,23,24}.\\
	
    Among the most remarkable phenomena observed in cavity magnomechanical platforms is magnomechanically induced transparency (MMIT), which originates from destructive quantum interference between different excitation pathways mediated by magnon–phonon coupling \cite{19,25,26}. Analogous to electromagnetically induced transparency in atomic systems \cite{27,28,29}, MMIT produces a narrow transparency window within an otherwise opaque spectral response, accompanied by steep dispersion and enhanced group delay \cite{30}. This effect provides an efficient mechanism for controlling microwave signal propagation, enabling applications such as slow and fast light manipulation \cite{32}, tunable microwave filtering \cite{33}, and coherent information storage \cite{34,35}. Furthermore, the strong tunability of coupling strengths and external magnetic fields in YIG-based systems allows flexible control over the MMIT characteristics, making cavity magnomechanical platforms highly promising for re-configurable quantum devices and hybrid microwave signal processing applications \cite{36,37,38}.\\
    
    On the other hand, nonreciprocity refers to the directional asymmetry in light propagation, whereby optical transmission is allowed in one direction while being suppressed in the opposite direction. Owing to this unidirectional transport property, nonreciprocal phenomena play a crucial role in quantum communication and the construction of complex quantum networks \cite{39,40}. In recent years, the Sagnac effect induced by spin resonators has been widely exploited to investigate a variety of nonreciprocal phenomena, including nonreciprocal entanglement and quantum steering \cite{41,42,43}, nonreciprocal photon blockade \cite{44}, and nonreciprocal squeezing \cite{45}. Meanwhile, substantial experimental progress has been achieved in the observation of Sagnac effect induced nonreciprocity \cite{46}. In addition, the magnon Kerr nonlinearity has been shown to induce nonreciprocal behavior by reversing the direction of the bias magnetic field \cite{47}, leading to the experimental realization of nonreciprocal tripartite entanglement \cite{48}. More recently, a scheme for realizing magnon blockade via the Barnett effect has been proposed \cite{49}, marking an important advance in nonreciprocal physics and opening new possibilities for exploring nonreciprocal entanglement between macroscopic quantum systems.\\
    
    Another important interference phenomenon closely related to induced transparency is the Fano resonance, which was first reported in atomic systems \cite{50}. It arises from quantum interference between different transition pathways, resulting in asymmetric spectral line shapes characterized by pronounced minima in the absorption profile \cite{51}. Subsequently, Fano resonances have been extensively investigated in a variety of physical platforms, including photonic crystals \cite{52}, coupled micro-resonators \cite{53}, and optomechanical systems \cite{54}. More recently, Fano-like asymmetric spectral features have been experimentally observed in hybrid cavity magnomechanical systems \cite{55}, highlighting the role of magnon–photon–phonon interference and further demonstrating the versatility of these platforms for controlling wave propagation and spectral responses.\\
    
    In this paper, we focus on the nonreciprocity of magnomechanically induced transparency, slow and fast light propagation, and Fano resonances arising from the Barnett effect in a cavity magnomechanical system. The Barnett effect has been extensively studied in ferromagnetic insulators, where mechanical rotation of a body carrying magnetic moments gives rise to induced magnetization \cite{56}, as well as in nuclear spin systems \cite{57}. This effect has enabled a range of applications, including remote control of magnetization switching \cite{58}, exploration of rotational vacuum friction \cite{59}, and detection of angular-momentum compensation points \cite{60}. In particular, the Barnett effect refers to the magnetization generated in a magnetic medium, such as a YIG sample, when it is subjected to rapid rotation. The resulting shift in the Barnett frequency can be continuously tuned to either positive or negative values by reversing the direction of the applied magnetic field, providing an efficient means to control magnonic and interference phenomena in hybrid magnomechanical platforms \cite{61,62,63}. Furthermore, we explore slow/fast light propagation, demonstrating that the group delay can be controlled by tuning the Barnett effect and photon-phonon interactions of the mirror within the YIG based system.\\
    
    The remainder of this paper is organized as follows. In Section \ref{01}, we introduce the system Hamiltonian, derive the corresponding quantum Langevin equations, and obtain the expression for the output field. Section \ref{0} is devoted to the investigation of magnomechanically induced transparency, with a particular focus on how different coupling forces and the Barnett effect influence the input-output spectrum. We also discuss the effect of the cavity decay rate and the dissipation rate of the two magnons on absorption. Section \ref{02} explores the group delay associated with the slow and fast propagation of light. In section \ref{03}, we study the nonreciprocal: absorption and group delay under different system parameters. In Section \ref{04}, we assess the experimental feasibility of the proposed scheme using currently accessible parameters and realistic configurations. Finally, Sec. \ref{05} summarizes our main results and conclusions.
		\begin{figure}[t]
		\centering
		\includegraphics[scale=0.35]{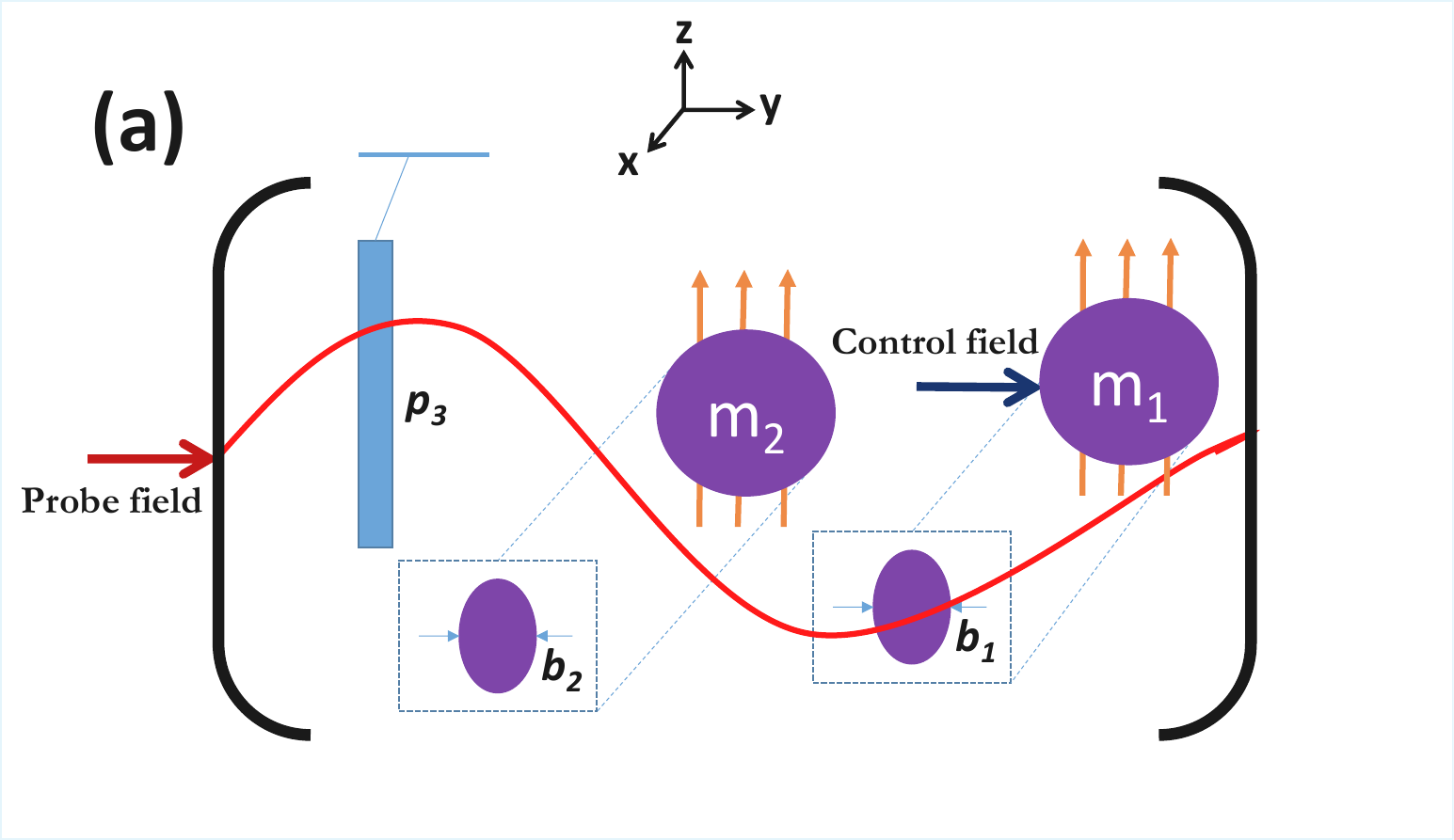}
	    \includegraphics[scale=0.34]{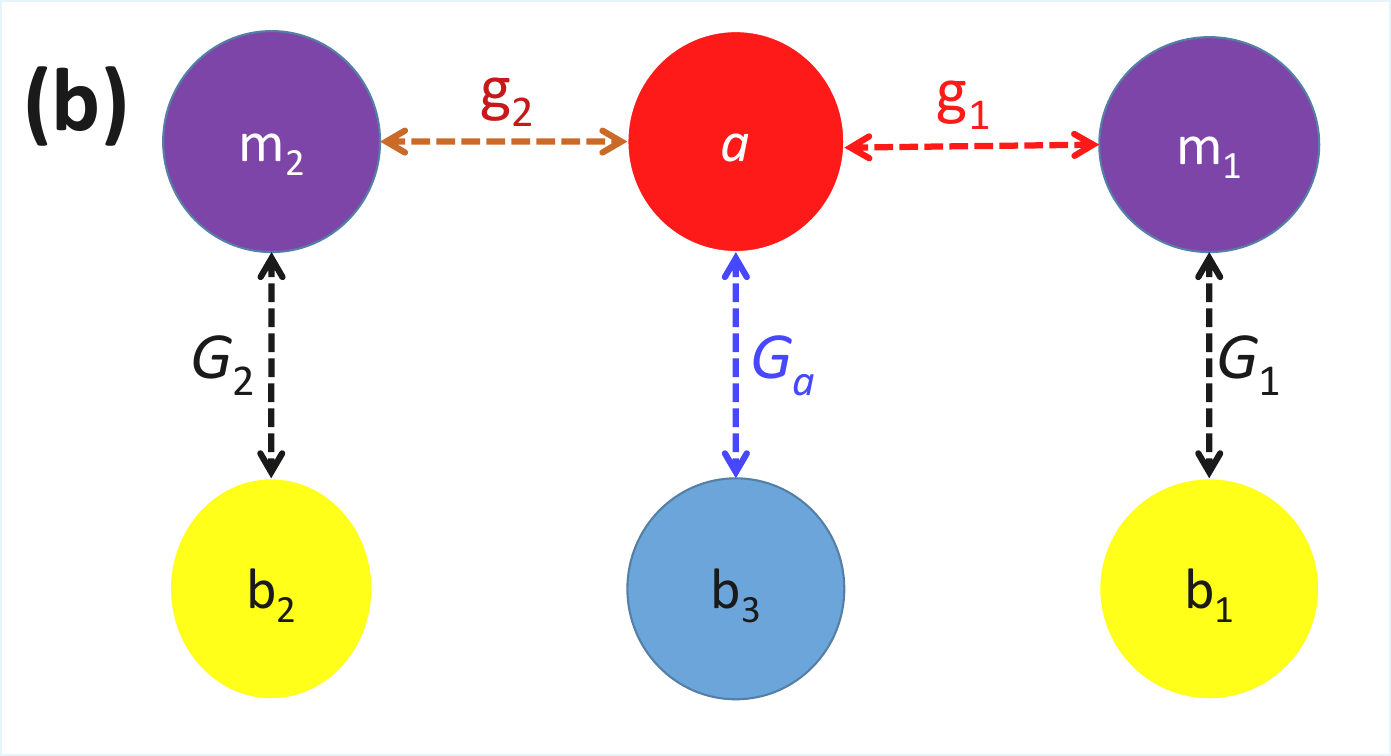}\\
	     \includegraphics[scale=0.35]{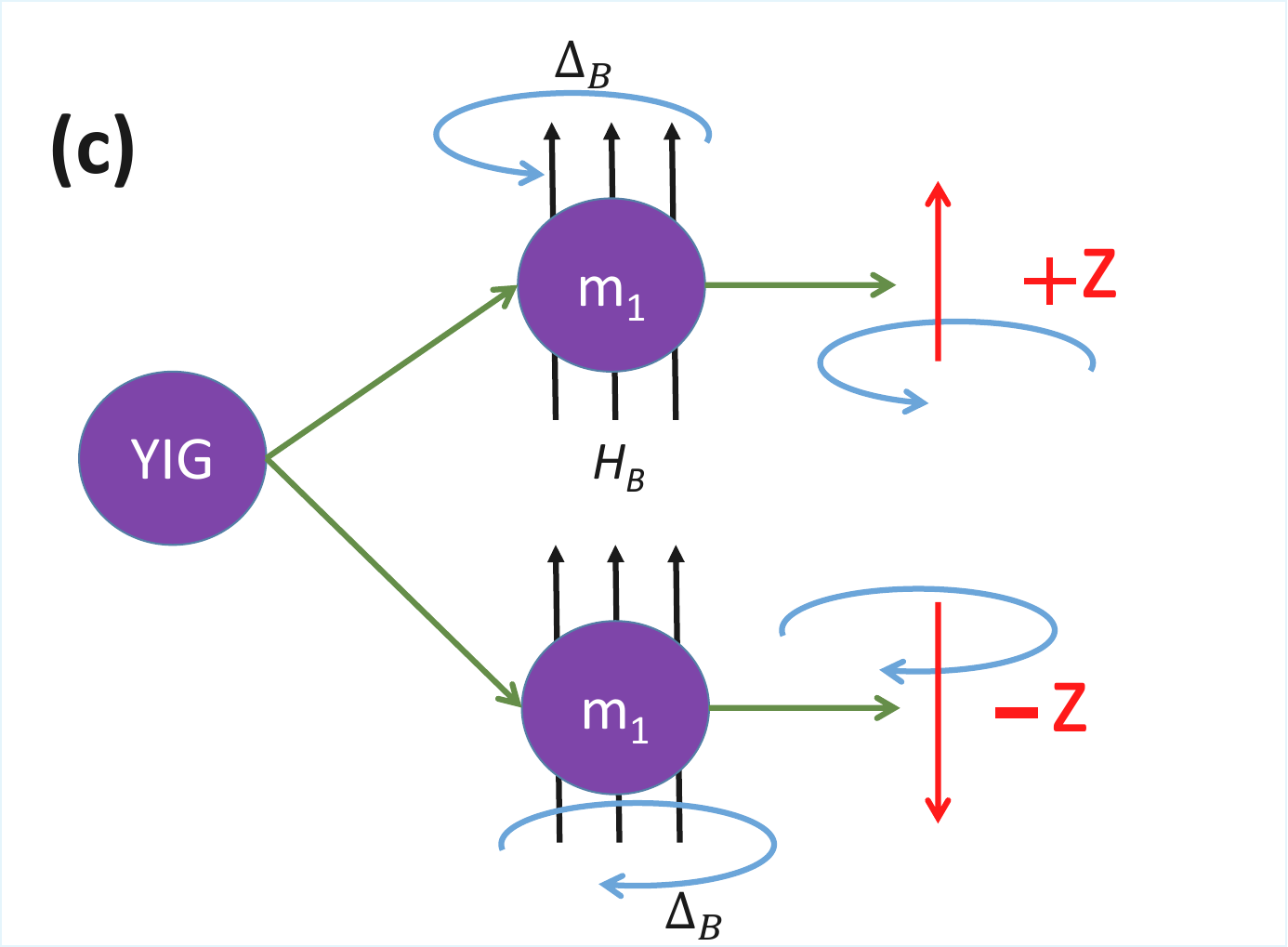}
	     \includegraphics[scale=0.38]{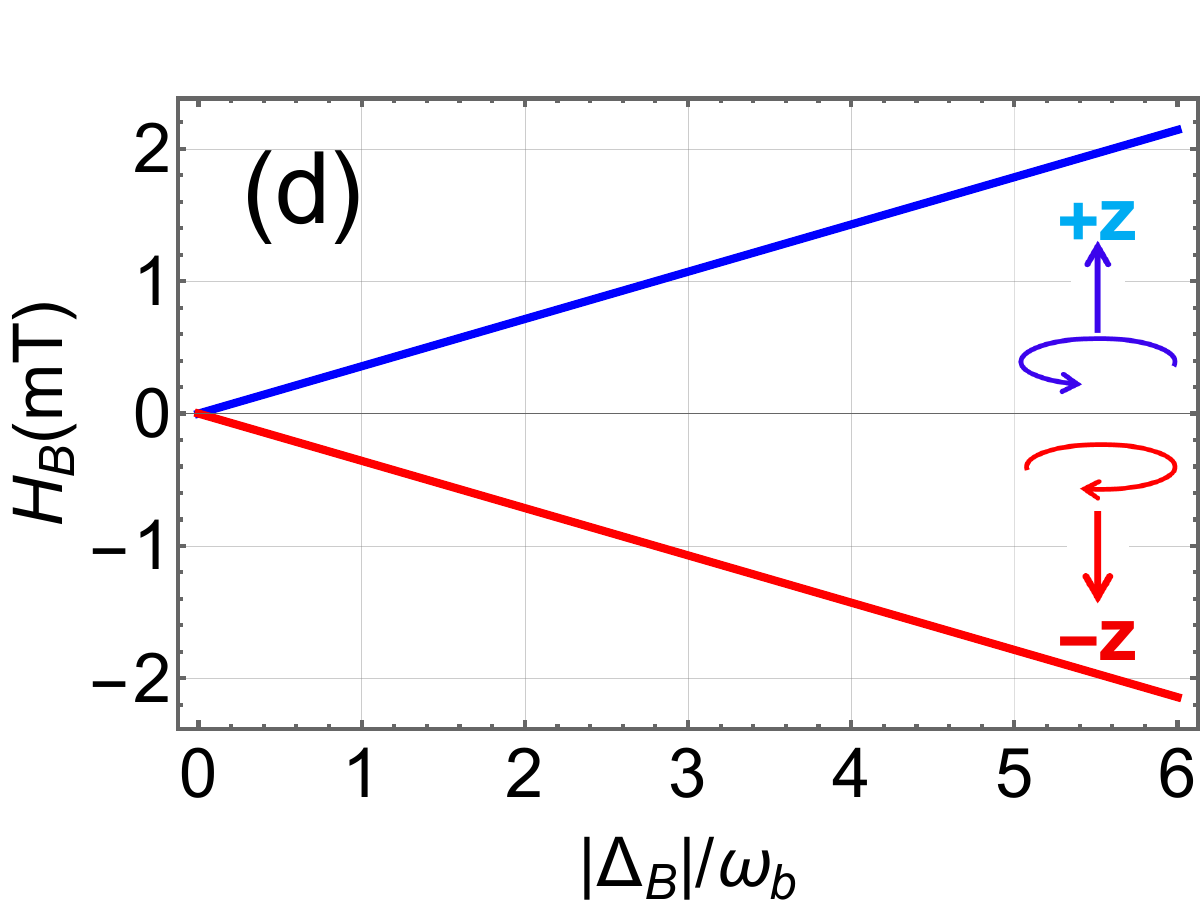}
		\caption{(a) Schematic representation of a hybrid cavity magnomechanical platform. The system comprises two ferromagnetic yttrium iron garnet (YIG) spheres and a mechanical membrane placed inside a microwave cavity, which is driven by an external probe field at frequency $\omega_{p}$. A uniform magnetic field applied along the z-axis excites the collective spin-wave (magnon) modes in each YIG sphere, which are strongly coupled to the cavity electromagnetic field. Furthermore, the bias magnetic field activates the magnetostrictive effect, thereby inducing magnon-phonon coupling within both YIG spheres and enabling coherent interactions among the cavity, magnon, and mechanical modes. However, the sphere $m_1$ is directly driven by an external microwave field applied along the y-direction (control field). The rotation of the YIG sphere at an angular frequency $\omega_{B}$ generates an effective magnetic field $H_{B}$, which induces a frequency shift in magnon $m_1$. (b) Diagram showing the couplings among the various modes of our system. (c) The insets show the rotation directions of the YIG sphere, which induce angular frequency shifts of $\pm\Delta_{B}$ through the Barnett effect. (d) Variation of $H_B$ as a function of $|\Delta_B|/\omega_{b}$ for different magnon rotation directions about the z-axis.}\label{fig1}
	\end{figure}
	\section{SYSTEM AND MODEL}\label{01}
  Fig.~\ref{fig1} illustrates a hybrid magnomechanical system composed of two ferromagnetic YIG spheres and a membrane positioned at the center of a microwave cavity. An external magnetic field applied along the \( z \)-axis excites the magnon modes in each sphere (\( m_1 \) and \( m_2 \)), which interact with the cavity field through magnetic dipole coupling. The mechanical vibrations of the spheres, induced by magnetostrictive forces, correspond to the phonon modes (\( b_1 \) and \( b_2 \)), thus enabling magnon--phonon coupling. In this configuration, a control field drives the first sphere (\( m_1 \)), while a probe field is injected into the cavity. The system is described by the following Hamiltonian
	\begin{equation}
		\begin{aligned}
			\mathcal{H}/\hbar = H_{{free}}+H_{{int}}+H_{{drive}},
		\end{aligned}
	\end{equation}
	where the free Hamiltonian terms are defined as
	\begin{equation}
	H_{{free}}=\omega_a a^{\dagger} a+\left(\omega_{m_1}+\Delta_B\right) m_1^{\dagger} m_1+\omega_{m_2} m_2^{\dagger} m_2+\sum_{r=1,2,3}\frac{\omega_{b_r}}{2}\left(p_r^2+q_r^2\right),
	\end{equation} 
	The first, second, and third terms represent the energies of the cavity mode, and magnon modes, respectively. Here, $\omega_{a}$ denotes the resonance frequency of the cavity, while $\omega_{m_1}$ and $\omega_{m_2}$ represents the frequencies of each magnon mode. The magnon frequency depends on the applied external magnetic bias field $H_{1(2)}$ and the gyromagnetic ratio $\gamma$, defined as $\omega_{m_{1(2)}} = \gamma H_{1(2)}$, with $\gamma / 2\pi = 28~\mathrm{GHz/T}$. We denote by $\Delta_B$ the angular frequency of the YIG sphere rotating about the $z$-axis. Owing to the Barnett effect~\cite{56,57,58,59,60}, the rotation of the YIG sphere generates an effective magnetic field, leading to the relation $\Delta_B = \gamma H_B$. As a result, the magnon mode frequency $\omega_{m_1}$ is shifted to $\omega_{m_1} + \Delta_B$. The operators $a~(a^{\dagger})$ and $m_{1(2)}~\left(m_{1(2)}^{\dagger}\right)$ describe the annihilation (creation) of cavity and magnon excitations, respectively, and satisfy the commutation relations $\left[a, a^{\dagger}\right] = 1$ and $\left[m_{1(2)}, m_{1(2)}^{\dagger}\right] = 1$. The fourth term accounts for the energy of three mechanical vibration modes with frequencies $\omega_{b_r}$, where $q_r$ and $p_r$, denote the dimensionless position and momentum operators of the phonon mode, fulfilling the commutation relation $[q_r, p_r] = i$, with $r = 1, 2,3$. The corresponding interaction Hamiltonian terms are expressed as follows
	\begin{equation}
		H_{{int}}=\sum_{r=1,2}\left(g_r\left(a m_r^{\dagger}+a^{\dagger} m_r\right)+G_{0 r} m_r^{\dagger} m_r q_r\right) -g_aa^\dagger a q_3, 
	\end{equation}
	the first term represents the energy associated with the interaction between the optical mode and the magnon mode, where \( g_{r} \) denotes the linear magnon-phonon coupling rate. The second term corresponds to the energy of the interaction between the mechanical mode and the magnon mode, with \( G_{0r} \) denoting the bare magnomechanical coupling rate. The last term characterizes the coupling between the mechanical mode and the optical mode, quantified by the rate $g_a$. The drive Hamiltonian terms are given by
	\begin{equation}
		H_{{drive}}=i \epsilon_p\left(a^{\dagger} e^{-i \omega_p t}-\text { H.c }\right)+i \Omega\left(m_1^{\dagger} e^{-i \omega_L t}-\text { H.c }\right),
	\end{equation}
the first term describes the energy of the optical driving field, characterized by the frequency \(\omega_p\) and the amplitude 
\(\epsilon_p = \sqrt{2\kappa_a P / \hbar \omega_p}\), where \(P\) is the probe power and \(\kappa_a\) denotes the cavity decay rate. The second term corresponds to the energy of the strong microwave field, which acts as the control field in the system. The associated Rabi frequency is defined as $\Omega = \frac{\sqrt{5}}{4}\gamma \sqrt{N} B$, representing the strength of the driving field. Here, \(N = \nu \mathcal{V}\) denotes the total number of spins in the YIG sphere, where 
\(\nu = 4.22 \times 10^{27}\,\text{m}^{-3}\) is the spin density of YIG and \(\mathcal{V}\) is the volume of the sphere. 
The parameters \(B\) and \(\omega_{L}\) correspond to the amplitude and frequency of the driving magnetic field, respectively. In this study, we investigate high-quality yttrium iron garnet (YIG) spheres with a diameter of 250~$\mu\text{m}$, composed of ferric ions (Fe$^{3+}$) with a density of $\rho = 4.22 \times 10^{27} \, \text{m}^{-3}$. This configuration yields a total spin of $S = \tfrac{5}{3} \rho V = 7.07 \times 10^{14}$, where $S$ denotes the collective spin operator satisfying the commutation relation $[S_\alpha, S_\beta] = i \epsilon_{\alpha\beta\gamma} S_\gamma$. Accordingly, the total Hamiltonian can be expressed as
\begin{equation} \label{ham}
	\begin{aligned}
		\mathcal{H} / \hbar= & \omega_a a^{\dagger} a+\left(\omega_{m_1}+\Delta_B\right) m_1^{\dagger} m_1+\omega_{m_2} m_2^{\dagger} m_2+\sum_{r=1,2,3}\frac{\omega_{b_r}}{2}\left(p_r^2+q_r^2\right)+\sum_{r=1,2}\left(g_r\left(a m_r^{\dagger}+a^{\dagger} m_r\right)+G_{0 r} m_r^{\dagger} m_r q_r\right) \\
		&-g_aa^\dagger a q_3 +i \epsilon_p\left(a^{\dagger} e^{-i \omega_p t}-\text { H.c }\right)+i \Omega\left(m_1^{\dagger} e^{-i \omega_L t}-\text { H.c }\right)
	\end{aligned}
\end{equation}
From the Hamiltonian in Eq.~\eqref{ham}, the quantum Heisenberg–Langevin equations in the rotating frame at frequency $\omega_L$ are given by
	\begin{equation}\label{a1}
		\begin{aligned}
			&\begin{aligned}
				& \dot{a}=-(\kappa_a+i\Delta_a)a-i \sum_{r=1,2}g_rm_r+ig_aaq_3+\epsilon_p e^{-i\delta t}+\sqrt{2\kappa_a} a^{in}, \\
				& \dot{m_r}=-(\kappa_{m_r} +i \Delta_{m_r} )m_r-i g_ra-iG_{0r}m_rq_r+\Omega+\sqrt{2 \kappa_{m_r}} m_r^{i n}, \quad \quad r=1,2\\
				& \dot{p_r}=-\omega_{b_r} q_r-G_{0j}m_r^\dagger m_r-\gamma_jp_r+\xi_r, \quad \quad r=1,2\\
				& \dot{p_3}=- \omega_{b_3} q_3+g_a a^\dagger a-\gamma_3p_3+\xi_3, \\
			\end{aligned}\\
		\end{aligned}
	\end{equation}
	where $\delta = \omega_p - \omega_{L}$, $\Delta_a = \omega_a - \omega_{L}$, and $\Delta_{m_r} = \omega_{m_r} - \omega_{L}$ represent the respective detunings. The parameters $\kappa_{m_r}$ and $\gamma_{r}$ denote the dissipation rates of the magnon and mechanical modes, respectively. The operators $a^{\mathrm{in}}$ and $m_r^{\mathrm{in}}$ represent the input noise acting on the cavity and magnon modes, while $\xi_r$ denotes the Hermitian Brownian noise operator associated with the mechanical mode. Since the nonlinear quantum equations cannot, in general, be solved analytically, Eq.~\eqref{a1} can be addressed using the perturbation method. In this approach, the steady-state solutions are obtained to first order in $\epsilon_p$, such that
	\[
	\langle \mathcal{D} \rangle = \mathcal{D}_s + \mathcal{D}_{-} e^{-i \delta t} + \mathcal{D}_{+} e^{i \delta t},
	\]
	where $\mathcal{D} = \{ a, m_1, m_2, q_1, q_2, q_3, p_1, p_2,p_3\}$. Consequently, the steady-state solutions of the dynamical operators are given by
		\begin{equation}
		\begin{aligned}
			&  a_{s}=\frac{-ig_1m_{1s}-ig_2m_{2s}}{\kappa_a+i\tilde{\Delta}_a},\quad m_{1s}=\frac{\Omega-i g_1 a_{s}}{\kappa_{m1}+i \tilde{\Delta}_{m1}}, \quad m_{2s}=\frac{-i g_{2} a_{s}}{\kappa_{m2}+i \tilde{\Delta}_{m2}}, \\
			& q_{1s}=\frac{-G_{0 1}|m_{1s}|^2}{\omega_{b_1}}, \quad q_{2s}=\frac{-G_{0 2}|m_{2s}|^2}{\omega_{b_2}}, \quad q_{3s}=\frac{g_a|a_{s}|^2}{\omega_{b_3}}.
		\end{aligned}
	\end{equation}
	The solution for the cavity mode is written as (see Appendix \ref{AP})
	\begin{equation}
		a_-=\epsilon_p\times\left(h_1+\frac{ig_1M}{L}+\frac{ig_2X_8}{X_7}-\frac{G_{aa}Z_4}{Z_3}\right)^{-1}.
	\end{equation}	
  Using the standard input–output relation, $\epsilon_{out}=\epsilon_{int}-2\kappa_a\langle a\rangle$, the amplitude of the output field can be expressed as follows
  	\begin{equation} \label{cp}
  	\epsilon_{\text{out}}=\frac{2\kappa_aa_-}{\epsilon_p}=\epsilon_R+i\epsilon_I.
  \end{equation}
   The real part $\epsilon_R$ of Eq.~\eqref{cp} characterizes the absorption spectrum, while the imaginary part $\epsilon_I$ corresponds to the dispersion spectrum of the output field at the probe frequency.
	\section{Magnomechanically induced transparency and Fano resonances}\label{0}
In this section, we conduct a detailed analysis of the magnomechanically induced transparency in our hybrid system by varying the coupling strengths $g_1$, $g_2$, $G_1$, $G_2$, and $G_a$, along with the Barnett effect. Furthermore, we examine how variations in both the cavity decay rate and the magnon dissipation rate influence the emergence, width, and depth of the resulting transparency windows.
Our analysis is based on the effective parameters reported in a recent cavity magnomechanics experiment (Table~\ref{Tab})~\cite{0,19,26}.
\begin{table}[t]
	\caption{Parameters used in numerical simulation}\label{Tab} 
	\begin{tabular*}{\textwidth}{@{\extracolsep{\fill}} l c c c}
		\hline\hline
		Parameter & Symbol & Value & Unit \\
		\hline\hline
		Microwave mode frequency & $ \omega_a$ & $2\pi\times10^{10}$ & Hz \\
		Driving magnetic field frequency & $ \omega_L$ & $2\pi\times10^{10}$ & Hz \\
		Phonon mode frequency of magnon $m_1$ & $\omega_{b_1}$ & $ 2\pi\times10^{7}$ & Hz \\
		Phonon mode frequency of magnon $m_2$ & $\omega_{b_2}$ & $ 2\pi\times10^{7}$ & Hz \\
		Phonon mode frequency of central mirror & $\omega_{b_3}$ & $ 2\pi\times10^{7}$ & Hz \\
		Decay rate of cavity mode & $\kappa_a$ & $2\pi\times10^{6}$ & Hz \\
		Dissipation rate of phonon mode $b_1$ & $\gamma_1$ & $2\pi\times10^{2}$ & Hz \\
		Dissipation rate of phonon mode $b_2$ & $\gamma_2$ & $2\pi\times10^{2}$ & Hz \\
		Dissipation rate of phonon mode $b_3$ & $\gamma_3$ & $2\pi\times10^{2}$ & Hz \\
		Dissipation rate of magnon mode $m_1$ & $\kappa_{m_1}$ & $2\pi\times0.1\times10^{6} $ & Hz \\
		Dissipation rate of magnon mode $m_2$ & $\kappa_{m_2}$ & $2\pi\times0.1\times10^{6} $ & Hz \\
		Coupling rate magnon(1)-cavity mode & $g_1$ & $2\pi\times1.5\times 10^{6}$ & Hz \\
		Coupling rate magnon(2)-cavity mode & $g_2$ & $2\pi\times1.5\times 10^{6}$ & Hz \\
		Coupling rate cavity-phonon $b_3$ mode & $G_a$ & $2\pi\times2.5\times 10^{6}$ & Hz \\
		Coupling rate magnon(1)-phonon $b_1$ mode & $G_1$ & $2\pi\times1.5\times 10^{6}$ & Hz \\
		Coupling rate magnon(2)-phonon $b_2$ mode & $G_2$ & $2\pi\times3.5\times 10^{6}$ & Hz \\
		Detunings & $\Delta_a$, $\Delta_{m_1}$ $\Delta_{m_2}$ & $2\pi\times10^{7}$ & Hz \\
		\hline\hline
	\end{tabular*}
\end{table}
The driving magnetic fields have an amplitude $B_{1(2)} = 3.6\times10^{-5}~\text{T}$, corresponding to a drive power $P_{1(2)} = 7.6~\text{mW}$ and a coupling strength $G_{01(02)}/2\pi = 0.2~\text{Hz}$ for a YIG sphere of diameter $250~\mu\text{m}$. Under these conditions, the number of spins is $N \simeq 3.5\times10^{16}$, giving $|\langle m_r \rangle| \simeq 1.1\times10^{7}$ ($r=1,2$) and $\langle m_r^{\dagger} m_r \rangle \simeq 1.2\times10^{14} \ll 5N = 1.8\times10^{17}$, confirming that the linearization condition is well satisfied \cite{19}.\\
		\begin{figure} [h!] 
		\begin{center}
			\includegraphics[scale=0.25]{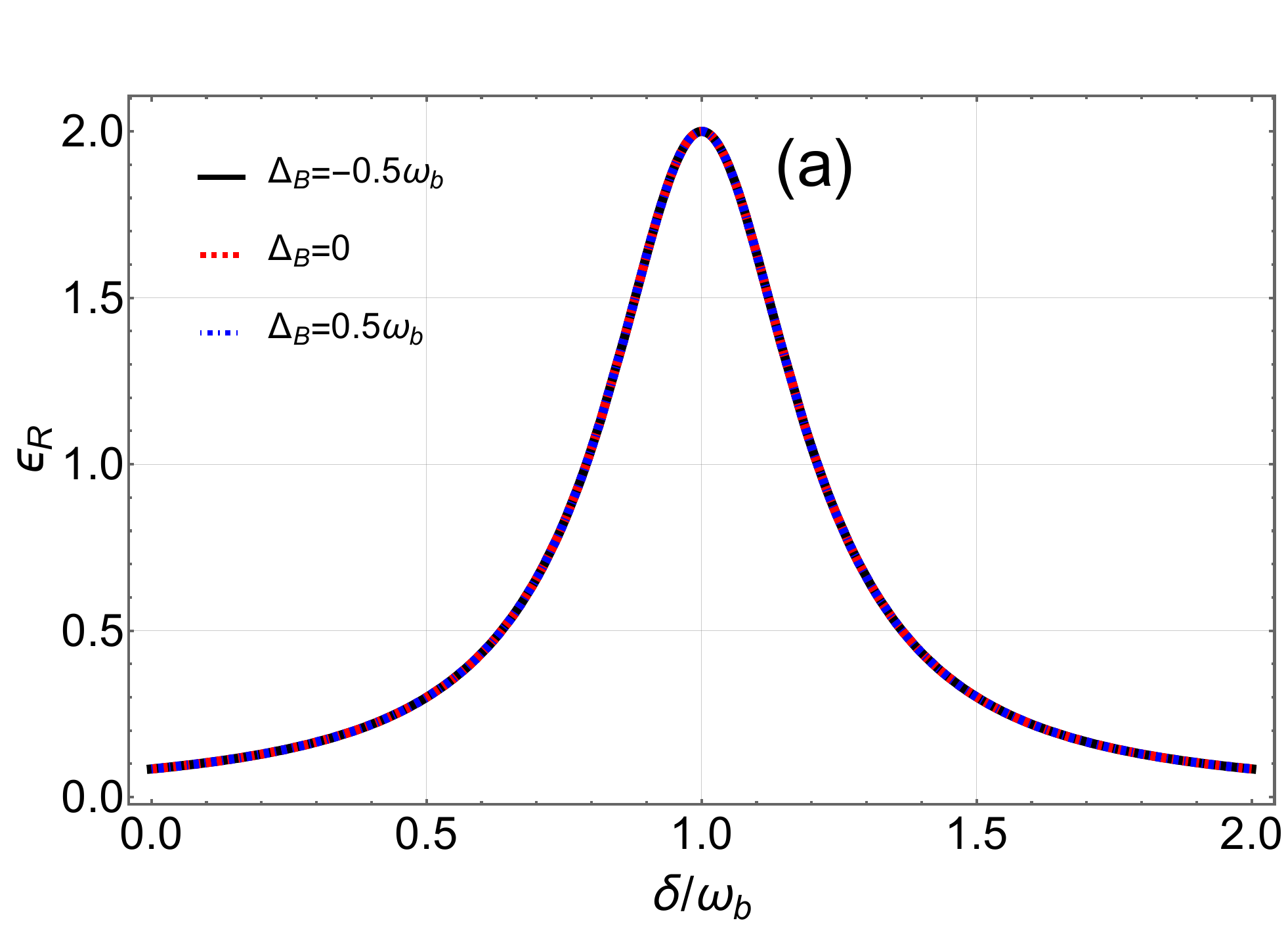}
			\includegraphics[scale=0.25]{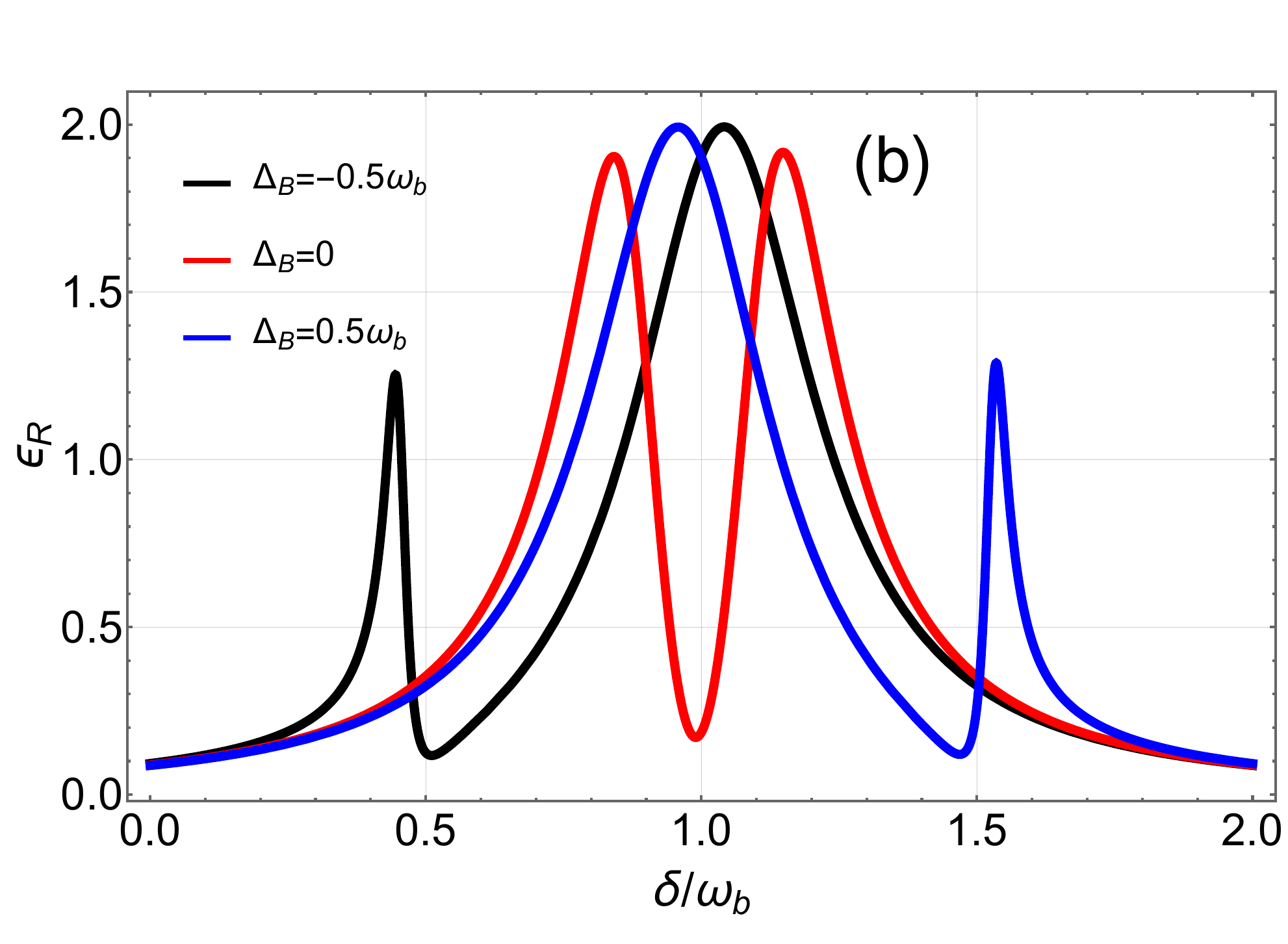}
			\includegraphics[scale=0.25]{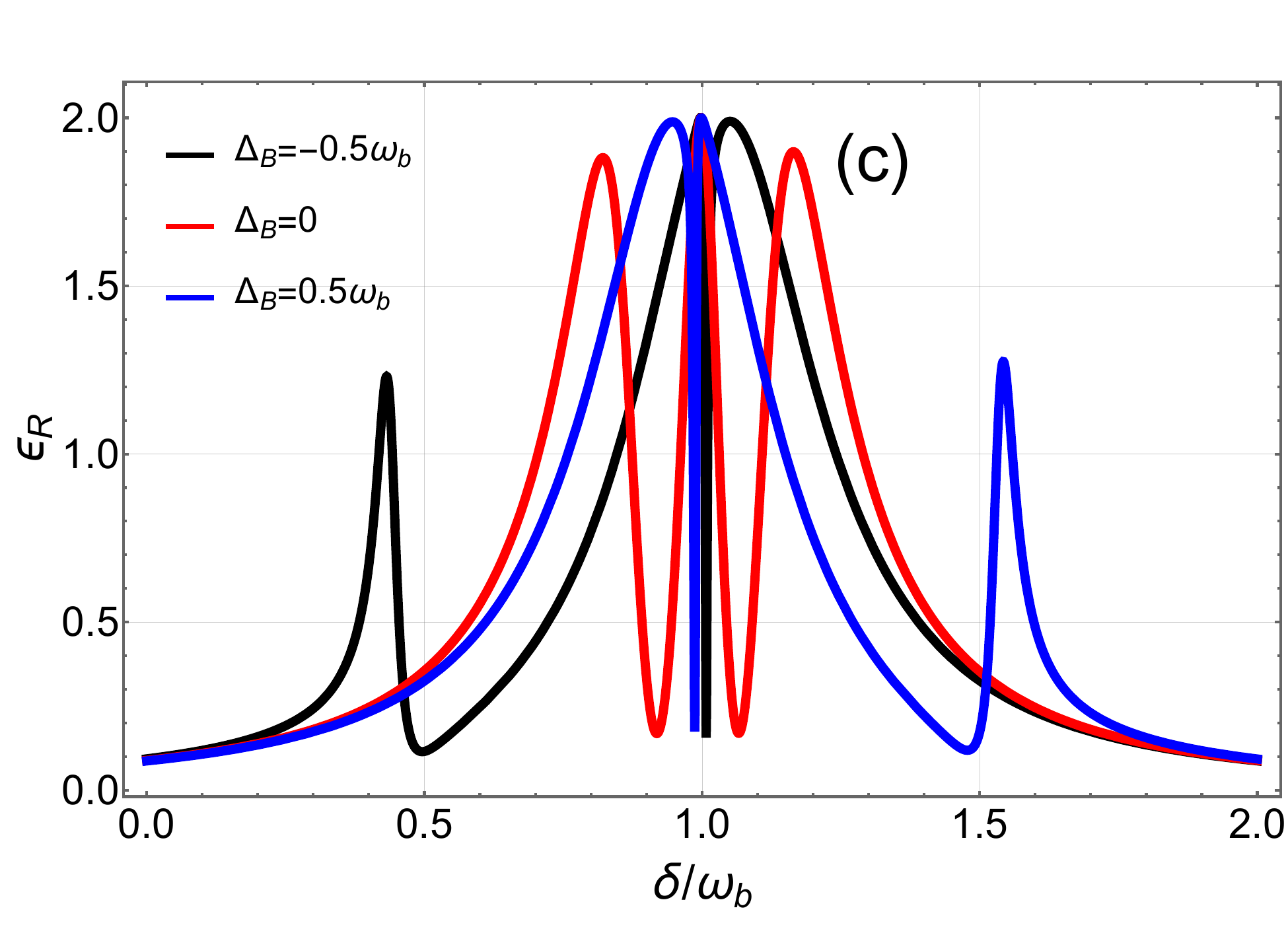}
			\includegraphics[scale=0.25]{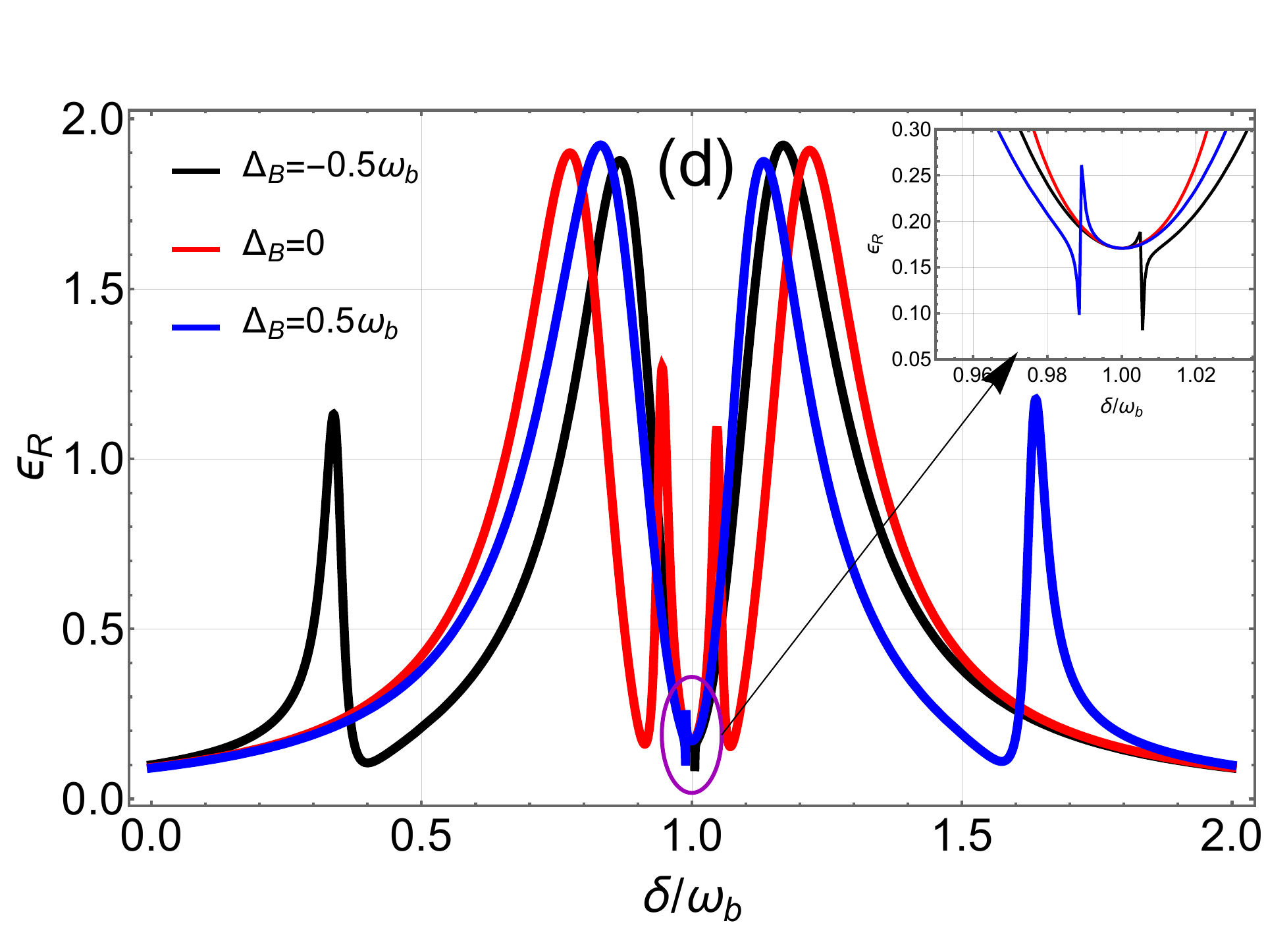}
			\includegraphics[scale=0.25]{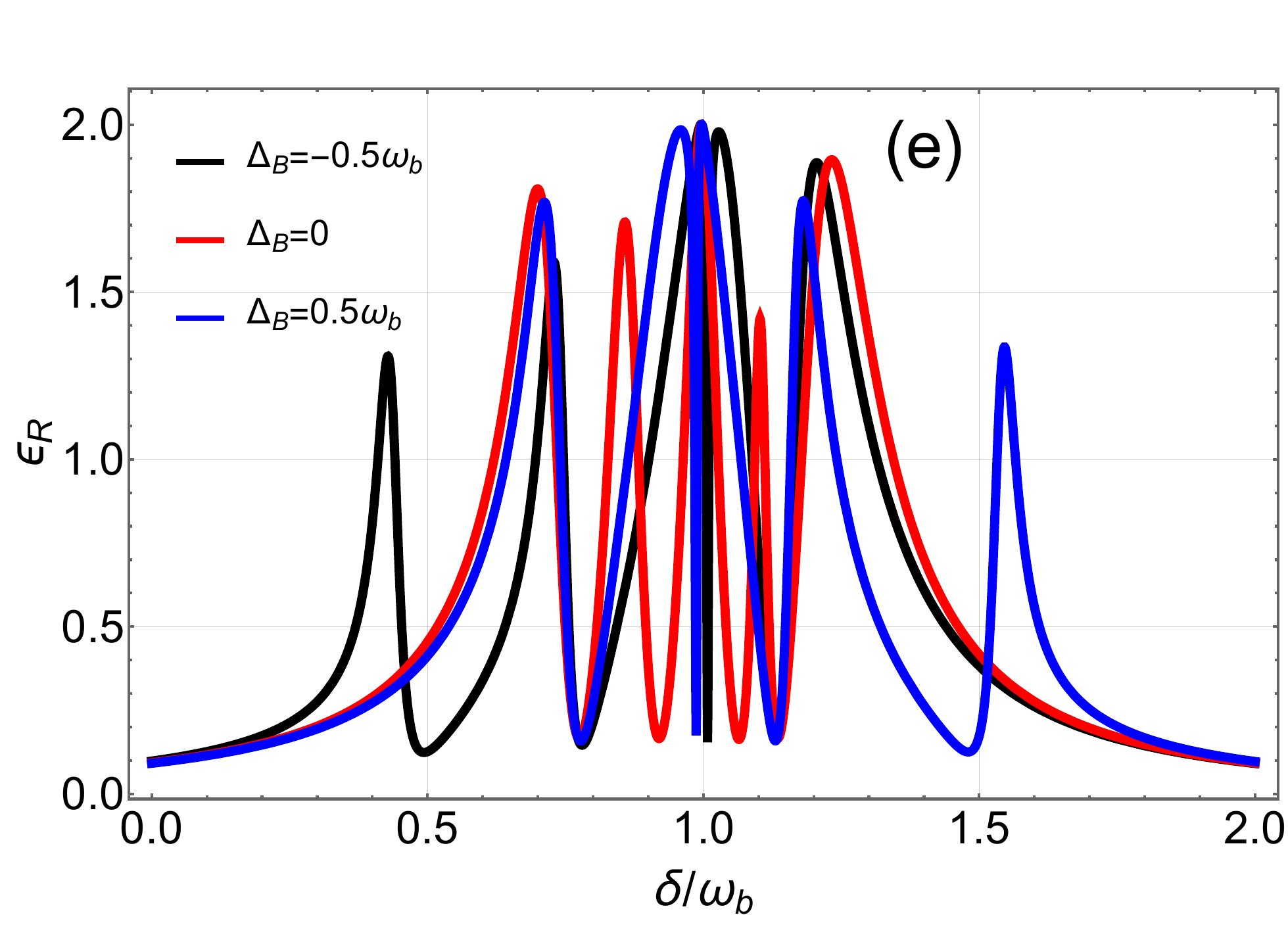}
			\includegraphics[scale=0.25]{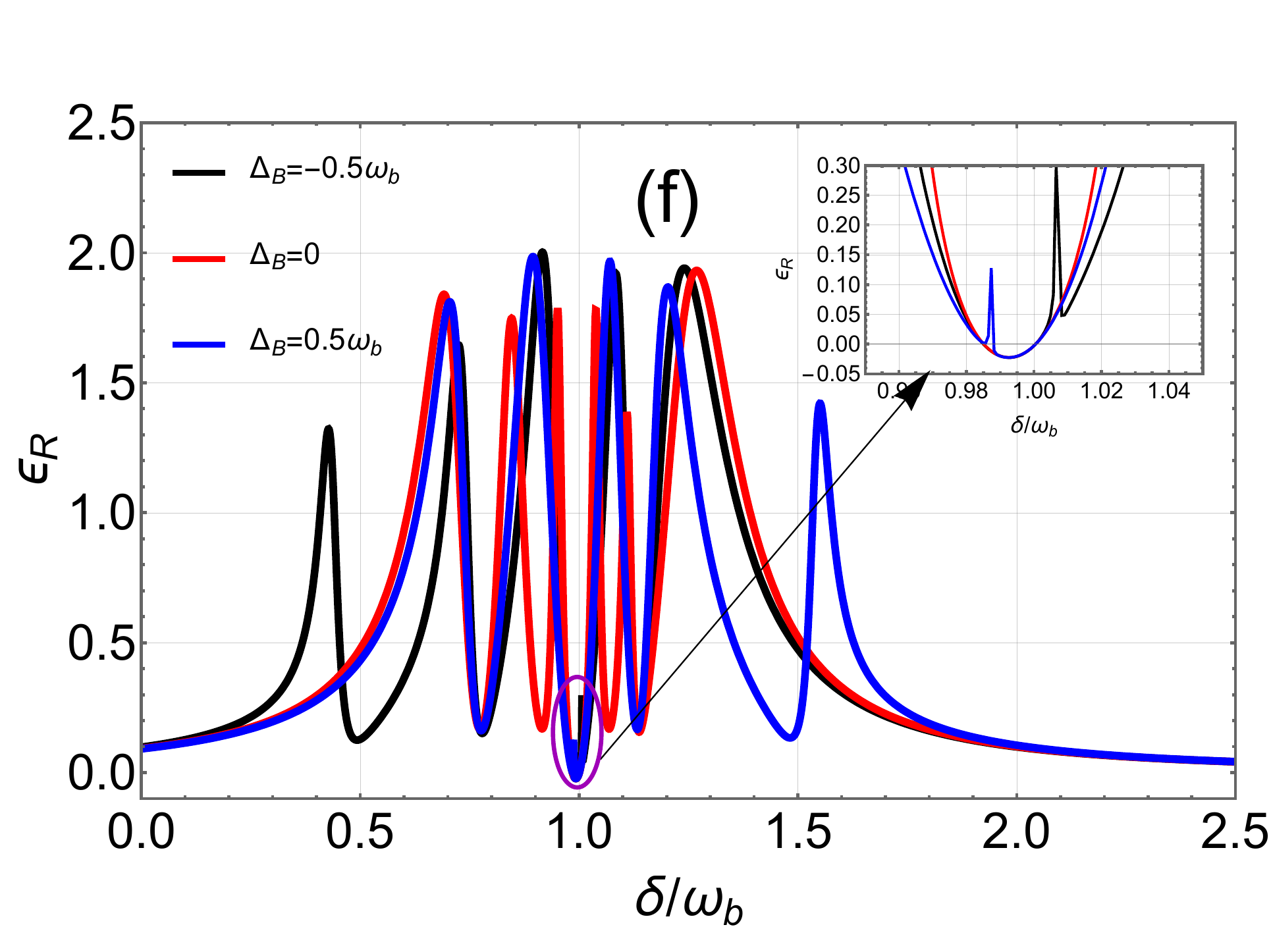}
			\caption{Real part of the output field $\epsilon_{\mathrm{R}}$ as a function of $\delta/\omega_{b}$. (a) $g_{1}=g_{2}=G_{1}=G_{2}=G_a=0$; (b) $g_{2}=G_{1}=G_{2}=G_a=0$ with $g_{1}/2\pi = 1.5$ MHz; (c) $g_{2}=G_{2}=G_a=0$ and $g_{1}/2\pi = G_{1}/2\pi = 1.5$ MHz;	(d) $G_{2}=G_a=0$ with $g_{1}/2\pi = g_{2}/2\pi = G_{1}/2\pi = 1.5$ MHz; (e) $G_a=0$ with $G_{2}/2\pi = 3.5$ MHz and $g_{1}/2\pi = g_{2}/2\pi = G_{1}/2\pi = 1.5$ MHz; (f) $G_{a}/2\pi = 2.5$ MHz with $g_{1}/2\pi = g_{2}/2\pi = G_{2}/2\pi = 1.5$ MHz and $G_{2}/2\pi = 3.5$ MHz. All remaining parameters are given in the Table~\ref{Tab}}\label{b}
		\end{center}
	\end{figure} 
	In Figs.~\ref{b}(a)--\ref{b}(f), we present the absorption spectrum of the output field as a function of the normalized probe detuning \( \delta / \omega_b \) for different coupling configurations. We first focus on the case where the Barnett effect is absent (\( \Delta_B = 0 \)), such that the observed spectral features originate solely from coherent interaction effects. As shown in Fig.~\ref{b}(a), the spectrum exhibits no indication of transparency windows when the photon mode is completely decoupled from the magnon modes ($g_{1}=0$ and $g_{2}=0$), the magnon mode is not coupled to the phonon modes ($G_{1}=0$ and $G_{2}=0$), and the photon-phonon interaction is switched off ($G_a=0$). Under these conditions, the system behaves as a purely absorbing medium, yielding a single broad absorption peak with no transparency features, regardless of the value of the Barnett effect (zero, positive, or negative). In Fig.~\ref{b}(b), only the first photon-magnon coupling $g_{1}$ is activated, while the second photon-magnon interaction ($g_{2}=0$), the magnon-phonon couplings ($G_{1}=0$ and $G_{2}=0$), and the photon-phonon coupling ($G_{a}=0$) remain switched off. In this configuration, the system begins to display a single transparency window in the absorption profile of the output probe field. This feature is evidenced by two symmetric absorption maxima separated by a distinct dip at the center, which reflects the emergence of destructive interference between the photon and magnon modes, an essential signature of induced transparency. In Fig.~\ref{b}(c), when the magnon-phonon coupling $G_{1}$ is activated while keeping $g_{2}=0$, $G_{2}=0$, and $G_{a}=0$, the absorption spectrum displays two distinct transparency windows. The photon-magnon interaction $g_{1}$, which initially produced a single MIT window in Fig.~\ref{b}(b), now leads to a splitting of this window into a double-transparency structure. The additional windows observed on the right side of Fig.~\ref{b}(c) originates from the magnon-phonon interaction $G_{1}$ and corresponds to the MMIT transparency window, reflecting the enhanced hybridization between the magnon and mechanical modes. In Fig.~\ref{b}(c), when the magnon-phonon coupling $G_{1}$ is switched on while keeping $g_{2}=0$, $G_{2}=0$ and $G_{a}=0$, the absorption spectrum exhibits two distinct transparency windows. The photon-magnon interaction $g_{1}$, which originally produced a single MIT window in Fig.~\ref{b}(b), now leads to a splitting of this window into two, as illustrated in Fig.~\ref{b}(d). The right-hand transparency feature observed in Fig.~\ref{b}(d) arises from the magnon-phonon coupling $g_{2}$ and corresponds to the magnomechanically induced transparency (MMIT) window, a phenomenon previously reported in Ref.~\cite{43}. Activating the magnon-phonon coupling $G_{2}$ while maintaining $G_{a}=0$ causes the transparency window to split into four distinct windows, producing a spectrum with five peaks and four troughs, as shown in Fig.~\ref{b}(e). Two of these windows result from magnon-phonon interaction, while the other two result from magnon-photon couplings. When all four coupling parameters are nonzero, i.e., $G_{a}\neq 0$, as shown in Fig.~\ref{b}(f), the transparency spectrum becomes richer and exhibits five distinct transparency windows. The central transparency window in Fig.~\ref{b}(f) is associated with the photon-phonon interaction, and corresponds to the so-called optomechanically induced transparency (OMIT) window. \\ 
	In the next, where we discuss the influence of the Barnett effect, we observe that when transparency windows are absent, the Barnett effect does not induce any noticeable change in the absorption spectrum. However, when transparency windows are present, the application of a positive Barnett-induced frequency shift causes the right transparency window to move toward higher detuning values, whereas a negative Barnett shift leads to a displacement of the left transparency window toward lower detuning values. As a consequence, the transparency profile becomes asymmetric. This asymmetry indicates the emergence of Fano-type resonance behavior in the transparency windows, originating from the interference between the discrete hybrid modes and the continuum background.\\
			\begin{figure} [h!] 
		\begin{center}
			\includegraphics[scale=0.4]{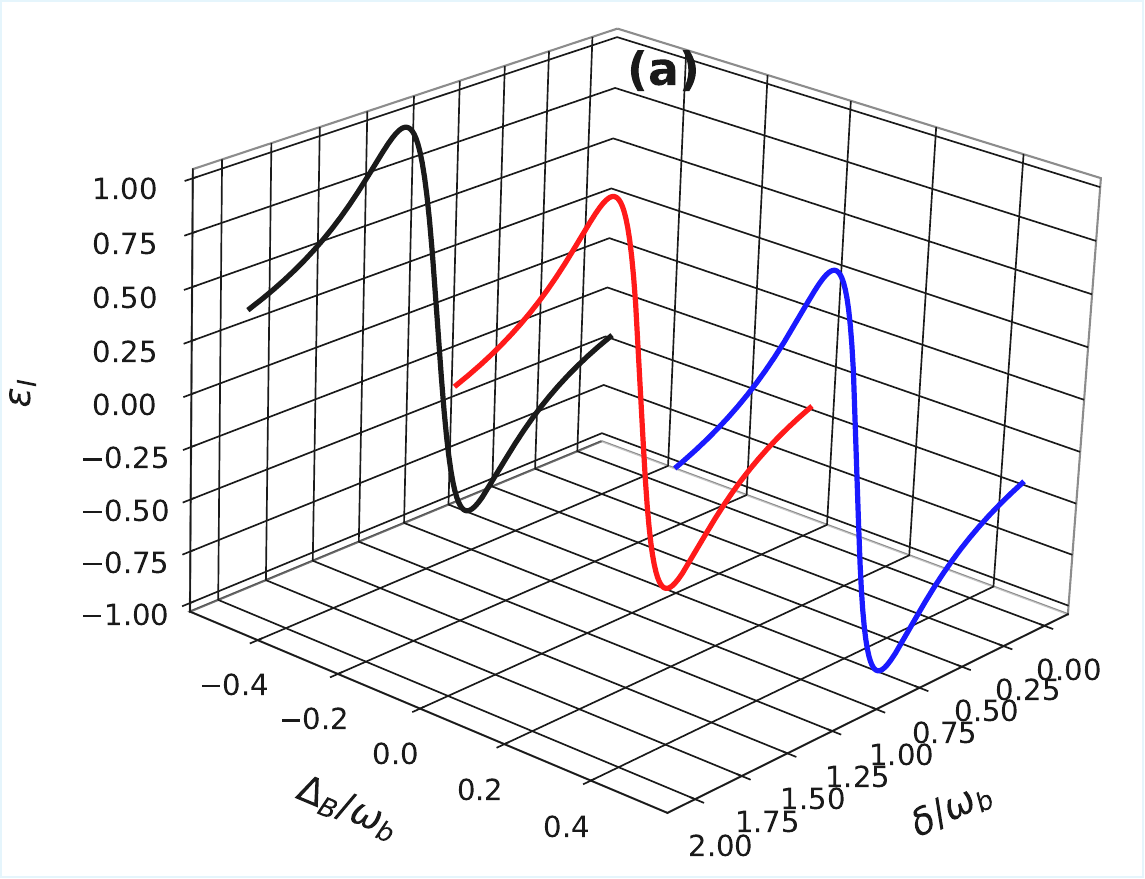}
			\includegraphics[scale=0.4]{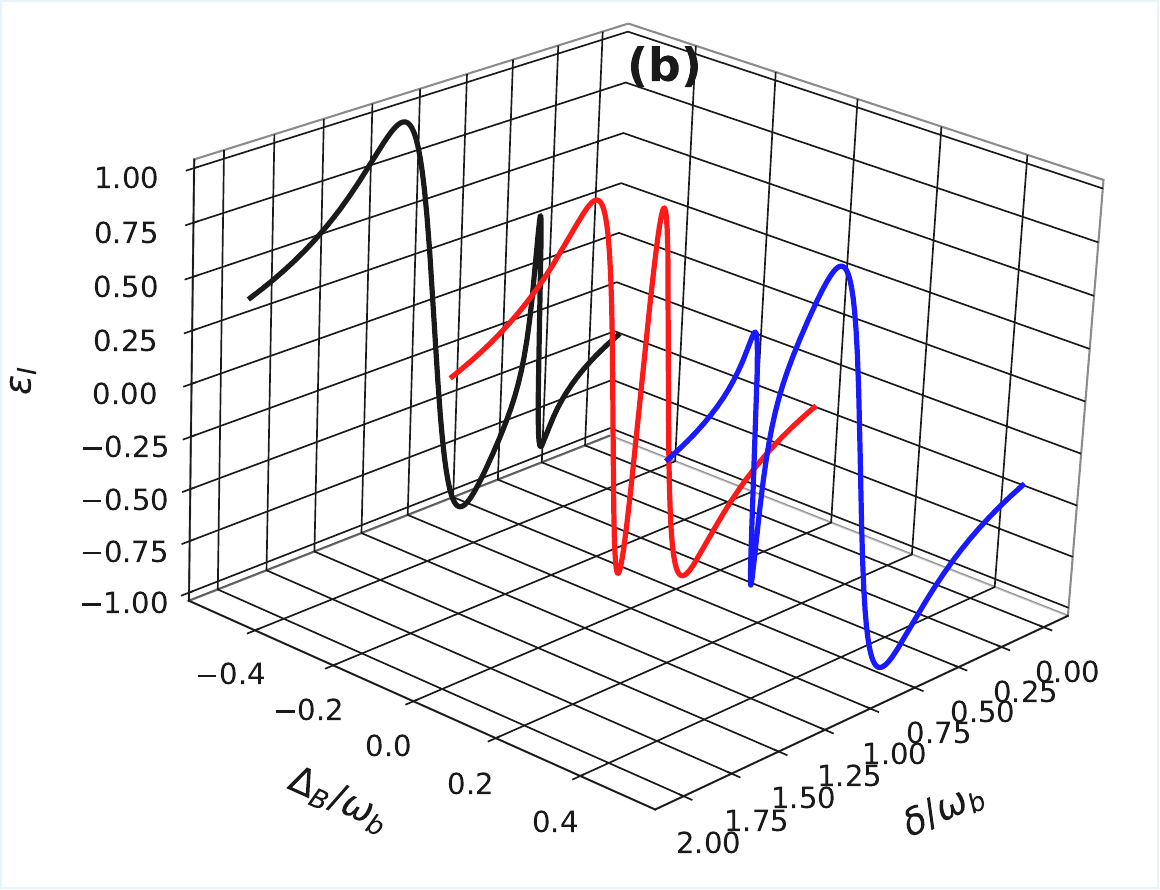}
			\includegraphics[scale=0.4]{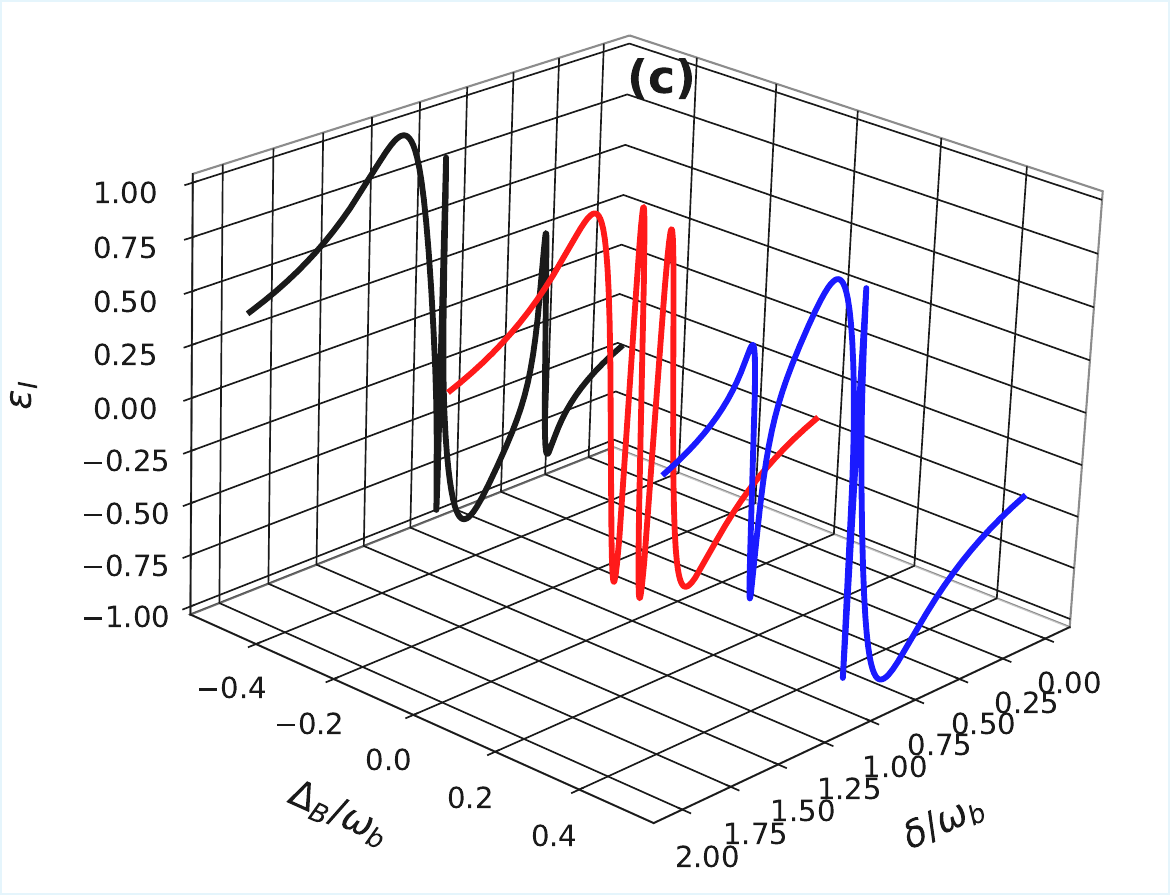}
			\includegraphics[scale=0.4]{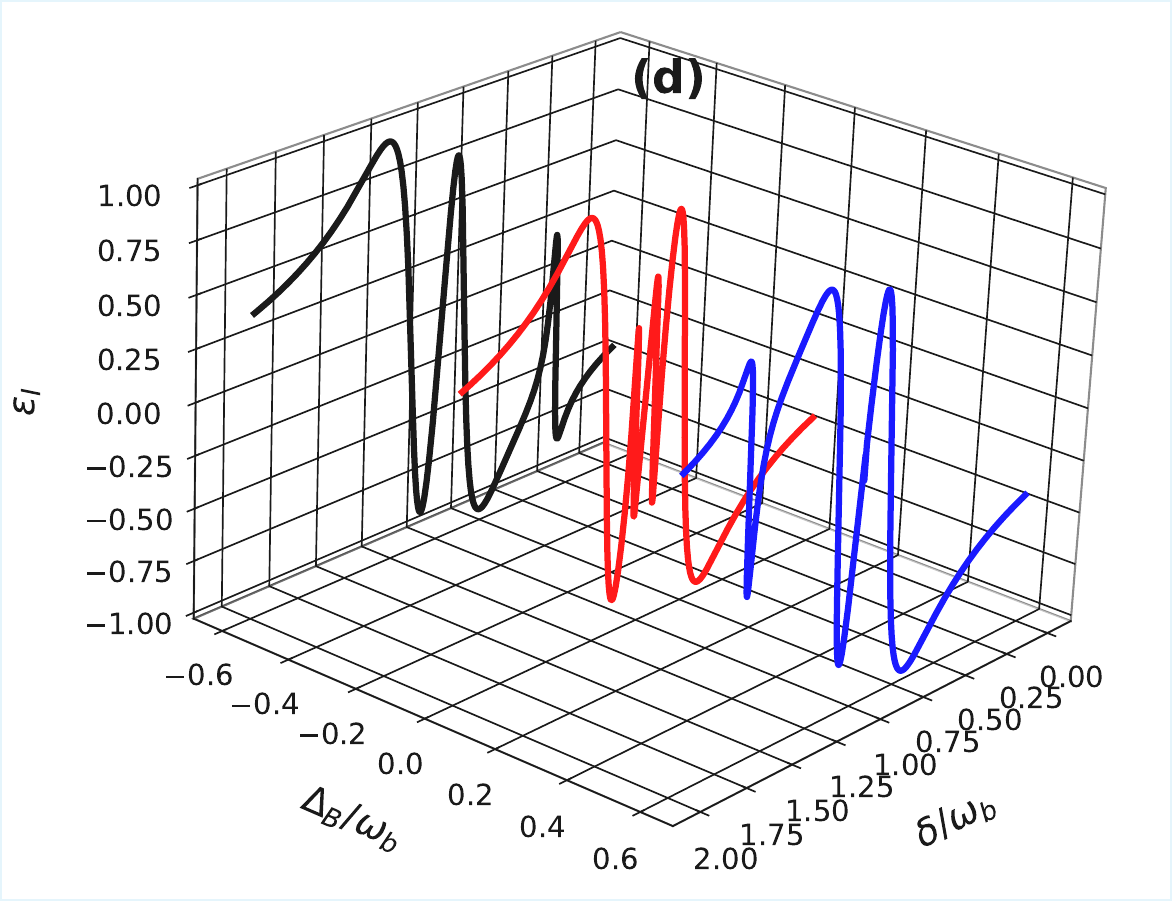}
			\includegraphics[scale=0.4]{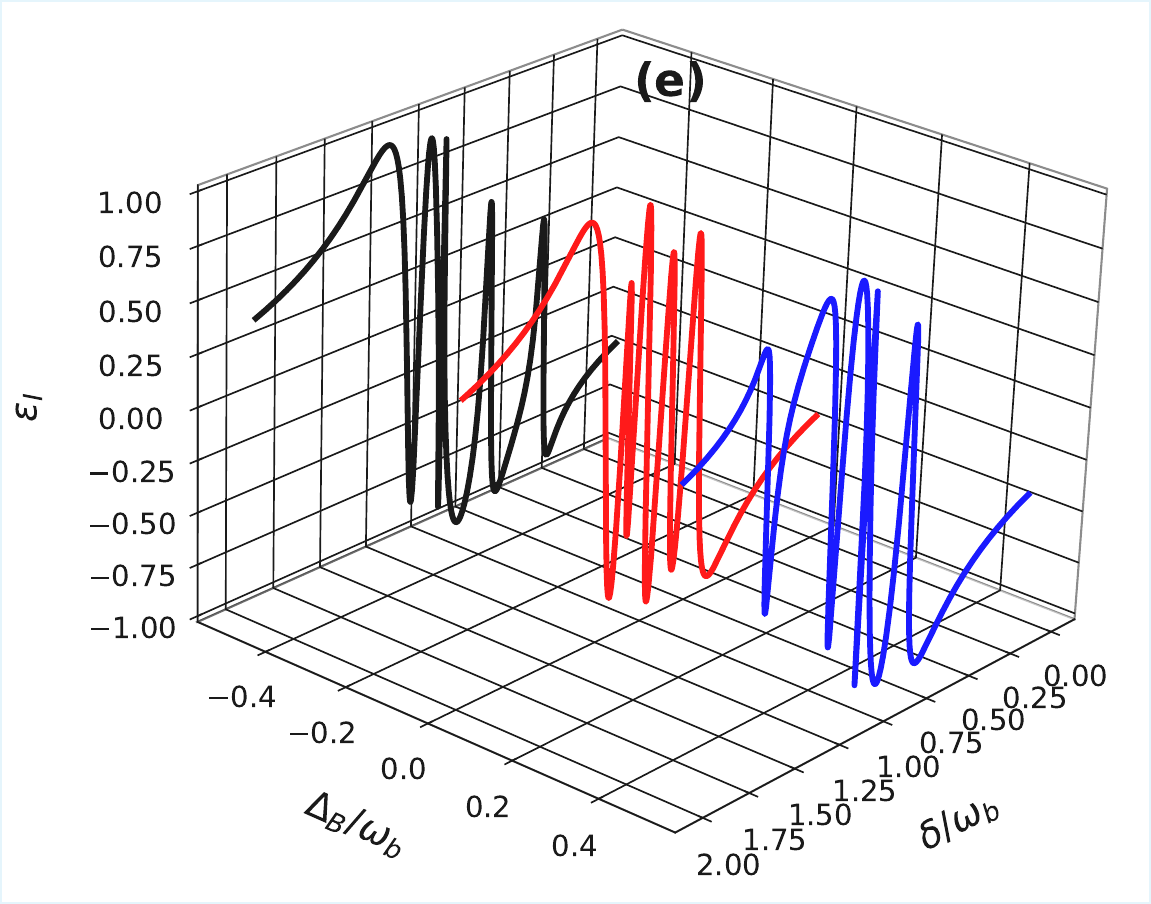}
			\includegraphics[scale=0.4]{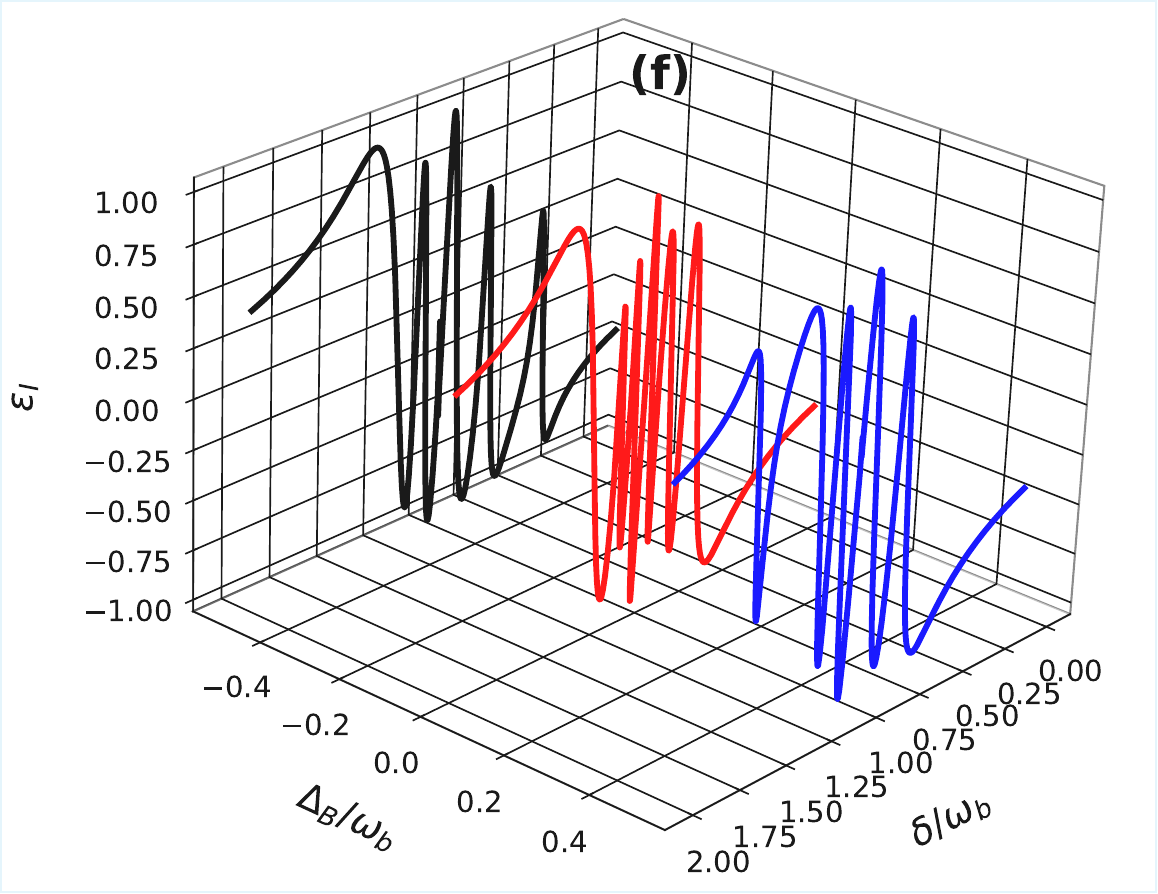}
			\caption{Imaginary part of the output field $\epsilon_{\mathrm{I}}$ as a function of $\delta/\omega_{b}$. (a) $g_{1}=g_{2}=G_{1}=G_{2}=G_a=0$; (b) $g_{2}=G_{1}=G_{2}=G_a=0$ with $g_{1}/2\pi = 1.5$ MHz; (c) $g_{2}=G_{2}=G_a=0$ and $g_{1}/2\pi = G_{1}/2\pi = 1.5$ MHz;	(d) $G_{2}=G_a=0$ with $g_{1}/2\pi = g_{2}/2\pi = G_{1}/2\pi = 1.5$ MHz; (e) $G_a=0$ with $G_{2}/2\pi = 3.5$ MHz and $g_{1}/2\pi = g_{2}/2\pi = G_{1}/2\pi = 1.5$ MHz; (f) $G_{a}/2\pi = 2.5$ MHz with $g_{1}/2\pi = g_{2}/2\pi = G_{2}/2\pi = 1.5$ MHz and $G_{2}/2\pi = 3.5$ MHz. All remaining parameters are given in the Table~\ref{Tab}.}\label{c}
		\end{center}
	\end{figure}
	In Figs.~\ref{c}(a)–\ref{c}(f), we present the dispersion spectrum of the transmitted field as a function of the scaled detuning $\delta/\omega_{b}$ for different coupling strengths and Barnett effect values. Fig~\ref{c}(a) shows that no signatures of MMIT appear in the dispersion spectrum when the magnon mode is decoupled from both the mechanical mode $(G_{1}=G_{2}=0)$ and the cavity mode $(g_{1}=g_{2}=0)$, and the photon–phonon coupling is also absent $(G_{a}=0)$. In Fig.~\ref{c}(b), activating the photon–magnon coupling $g_{1}$ while keeping $g_{2}=0$, $G_{1}=0$, $G_{2}=0$ and $G_{a}=0$ leads to the emergence of a single transparency window in the dispersion spectrum. In Fig.~\ref{c}(c), when the magnon–phonon coupling $G_{1}$ is introduced, the single transparency window splits into two distinct windows. When activate the photon-magnon coupling $g_2$, we observe three MMIT dispersion spectrum as shown in Fig.~\ref{c}(d). Figures~\ref{c}(e) and \ref{c}(f) reveal four and five transparency windows in the dispersion spectrum, respectively, once the additional magnon–phonon and photon–phonon couplings are switched on. Concerning the influence of the Barnett effect, Fig.~\ref{c}(a) remains unchanged since no MMIT features are present in the absence of coherent coupling induced interference. However, in Figs.~\ref{c}(b)–\ref{c}(f), where MMIT is clearly established, the Barnett effect induces an asymmetric frequency shift of the transparency windows. Specifically, for a negative Barnett detuning $(\Delta_{B}=-0.5\omega_{b})$, the left transparency windows are shifted toward lower detuning values, whereas for a positive Barnett detuning $(\Delta_{B}=0.5\omega_{b})$, the right transparency windows move toward higher detuning values. This behavior demonstrates that the Barnett effect provides an efficient tuning mechanism for controlling the spectral position and asymmetry of MMIT induced dispersion features.\\
	\begin{figure} [h!] 
		\begin{center}
			\includegraphics[scale=0.35]{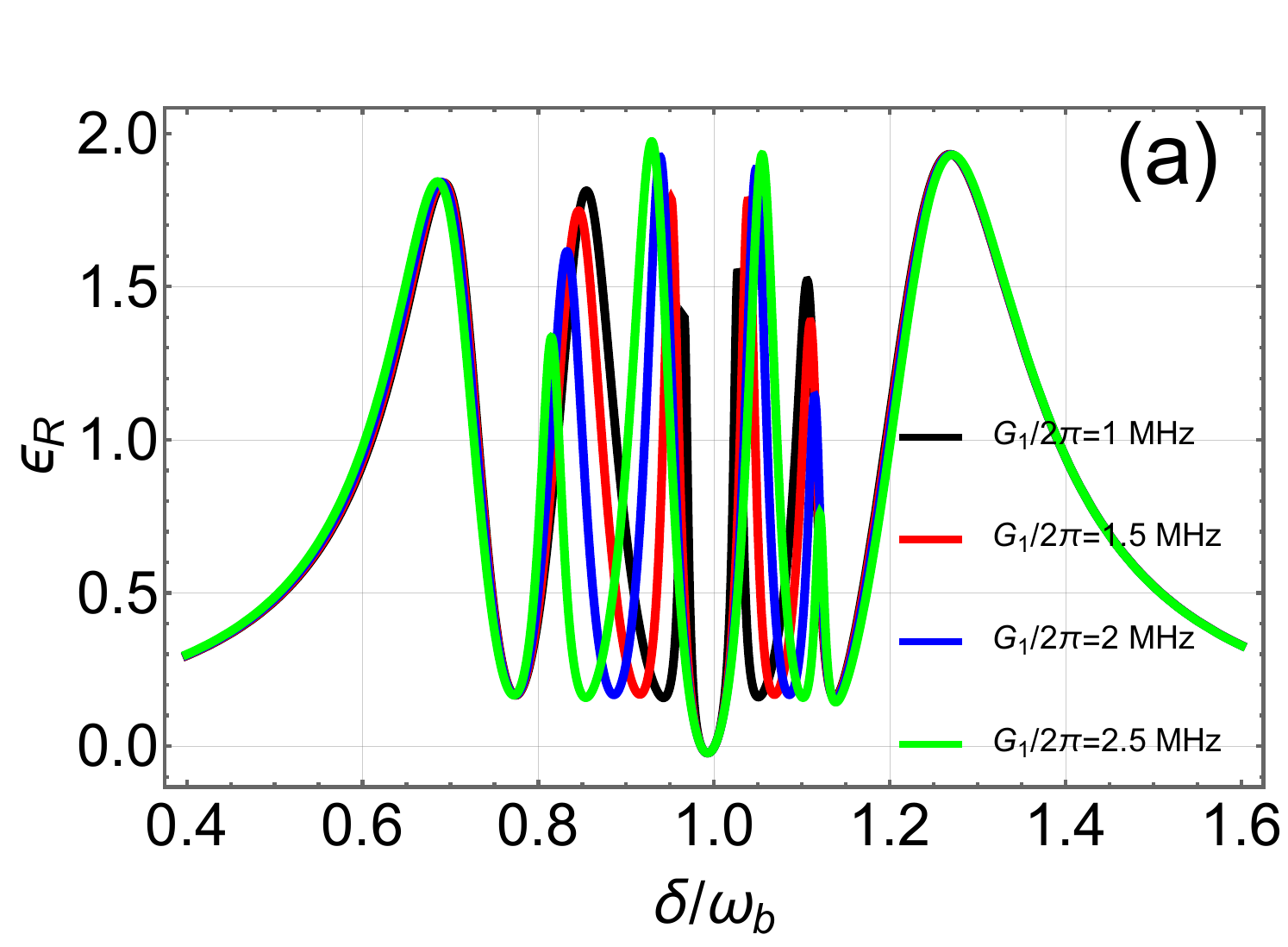}
			\includegraphics[scale=0.35]{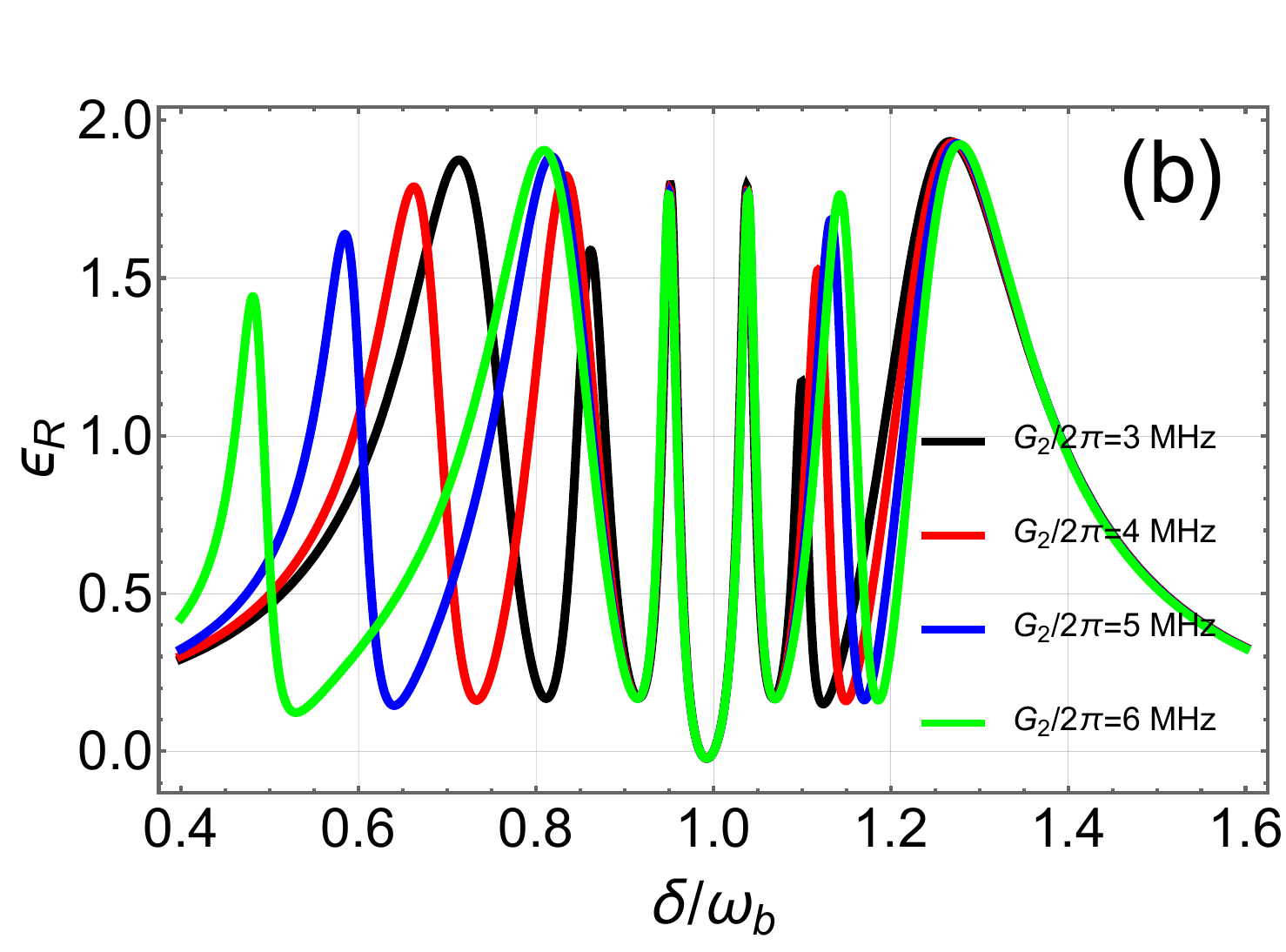}
			\includegraphics[scale=0.34]{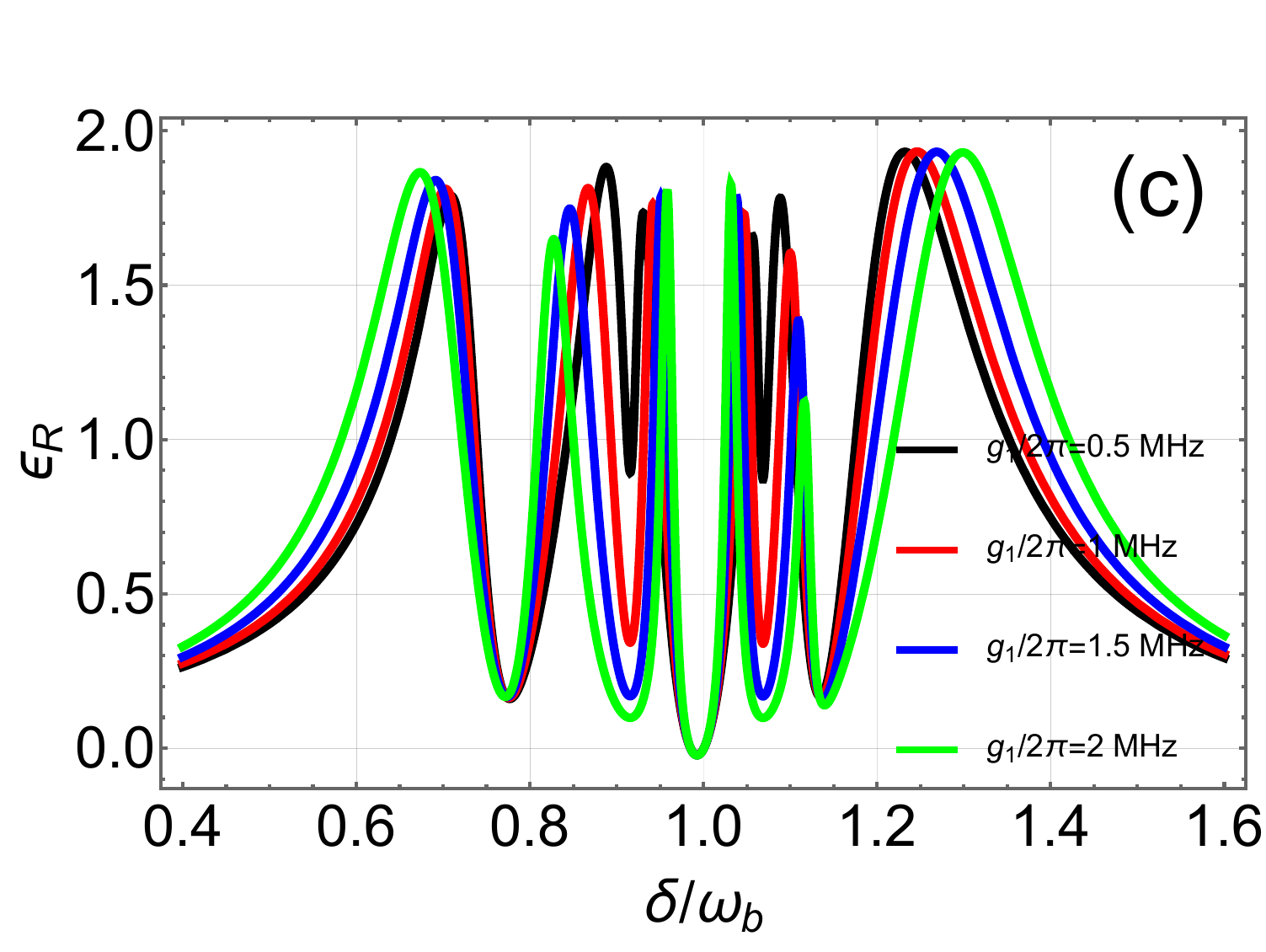}
			\includegraphics[scale=0.34]{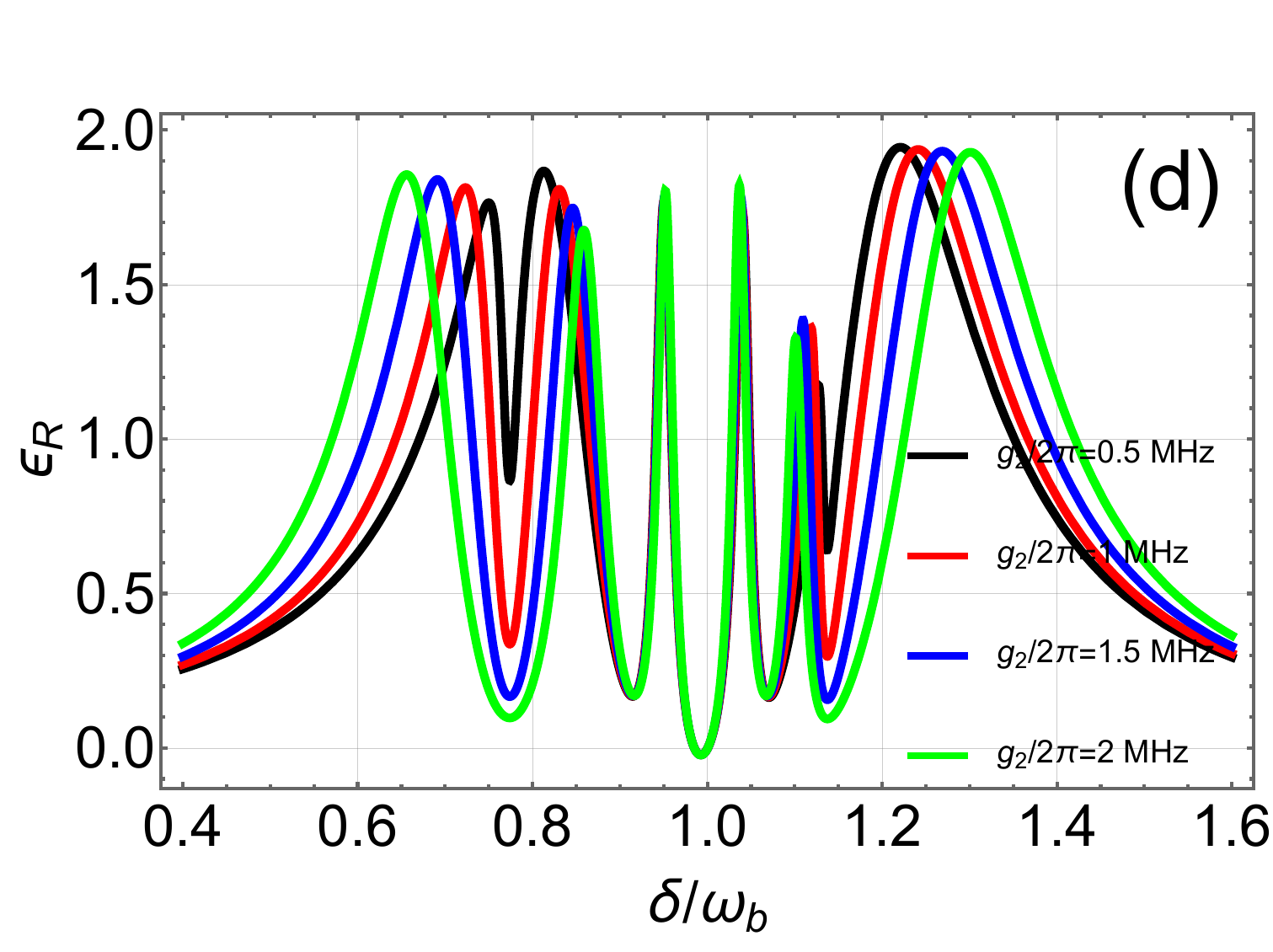}
			\includegraphics[scale=0.7]{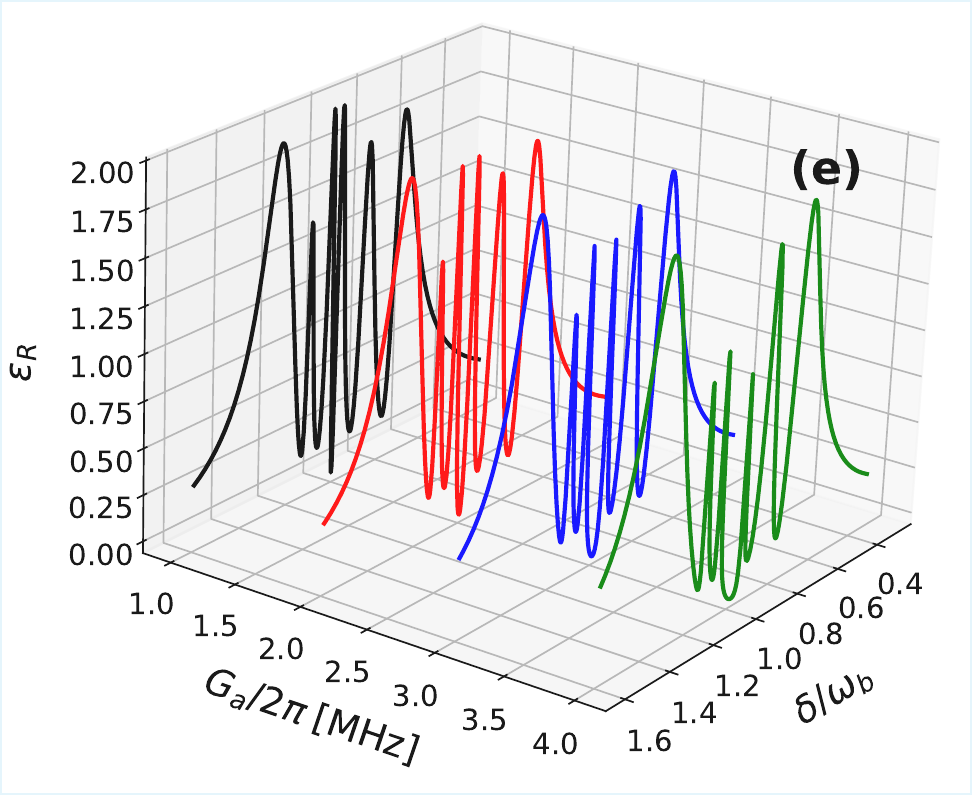}
			\caption{Real part of the output field $\epsilon_{\mathrm{R}}$ as a function of the normalized detuning $\delta/\omega_{b}$ for different coupling strengths: (a) varying $G_{1}$ with $g_{1}/2\pi = g_{2}/2\pi = 1.5~\text{MHz}$, $G_{2}/2\pi = 3.5~\text{MHz}$, $G_{a}/2\pi = 2.5~\text{MHz}$ and $\Delta_{B}=0$; (b) varying $G_{2}$ with $g_{1}/2\pi = g_{2}/2\pi = G_{1}/2\pi = 1.5~\text{MHz}$, $G_{a}/2\pi = 3~\text{MHz}$ and $\Delta_{B}=0$; (c) varying $g_{1}$ with $G_{1}/2\pi = g_{2}/2\pi = 1.5~\text{MHz}$, $G_{2}/2\pi = 3.5~\text{MHz}$, $G_{a}/2\pi = 2.5~\text{MHz}$ and $\Delta_{B}=0$; (d) varying $g_{2}$ with $G_{1}/2\pi = g_{1}/2\pi = 1.5~\text{MHz}$, $G_{2}/2\pi = 3.5~\text{MHz}$, $G_{a}/2\pi = 2.5~\text{MHz}$ and $\Delta_{B}=0$; and (e) varying $G_{a}$ with $g_{1}/2\pi = g_{2}/2\pi = G_{1}/2\pi = 1.5~\text{MHz}$, $G_{2}/2\pi = 3.5~\text{MHz}$ and $\Delta_{B}=0$. All remaining parameters are provided in the Table~\ref{Tab}.}\label{d}
		\end{center}
	\end{figure}
	 In Fig.~\ref{d}(a), we present the absorption spectrum $\epsilon_R$ of the output field for several values of the magnon–phonon coupling $G_{1}$, plotted as a function of the normalized probe detuning $\delta/\omega_{b}$. The parameters are fixed at $g_{1}/2\pi = g_{2}/2\pi = 1.5$ MHz, $G_{2}/2\pi = 3.5$ MHz, $G_{a}/2\pi = 2.5$ MHz and $\Delta_{B}=0$. For $G_{1}/2\pi = 1$ MHz, the dips of the first five transparency windows occur at $\delta = 0.77\omega_b$, $0.94\omega_b$, $0.99\omega_b$, $1.06\omega_b$, and $1.14\omega_b$, respectively. As $G_{1}$ increases, the second dip shifts to lower frequencies and the fourth dip moves to higher frequencies, while the remaining dips remain essentially unchanged. These features indicate that the second and fourth transparency windows are governed by the MIT mechanism and are strongly sensitive to variations in $G_{1}$. Fig.~\ref{d}(b) shows the corresponding behavior when varying the second magnon–phonon coupling $G_{2}$, with $g_{1}/2\pi = g_{2}/2\pi = G_{1}/2\pi = 1.5$ MHz, $G_{a}/2\pi = 2.5$ MHz and $\Delta_{B}=0$ held fixed. For $G_{2}/2\pi = 3$ MHz, the first five dips appear at $\delta = 0.81\omega_b$, $0.92\omega_b$, $0.99\omega_b$, $1.07\omega_b$, and $1.12\omega_b$, respectively. Increasing $G_{2}$ causes the first dip to shift to lower frequencies and the fifth dip to shift to higher frequencies, whereas the central three dips remain stationary. This behavior confirms that the first and fifth transparency windows are associated with the MIT effect and are selectively modified by the tuning of $G_{2}$. The influence of the magnon–photon coupling $g_{1}$ is illustrated in Fig.~\ref{d}(c). With $G_{1}/2\pi = g_{2}/2\pi = 1.5$ MHz, $G_{2}/2\pi = 3.5$ MHz, $G_{a}/2\pi = 3$ MHz, and $\Delta_{B}=0$ fixed, varying $g_{1}/2\pi$ from $0.5$ to $2$ MHz leads to a substantial broadening and deepening of the second and fourth transparency windows. These modifications reflect an enhanced interference pathway involving the first magnon mode, further confirming the MIT origin of these features. Fig.~\ref{d}(d) reports the corresponding behavior when varying the second magnon–photon coupling $g_{2}$, under the same fixed conditions for the remaining parameters. Increasing $g_{2}/2\pi$ from $0.5$ to $2$ MHz significantly strengthens the first and fifth transparency windows, while the central windows remain largely insensitive. This outcome mirrors the trends observed with $g_{1}$ but targets a different subset of transparency features, again linking the enhanced response to MIT processes mediated by the second magnon mode. Finally, Fig.~\ref{d}(e) examines the impact of the photon–phonon coupling $G_{a}$, with the magnon related interactions kept fixed at $g_{1}/2\pi = g_{2}/2\pi = G_{1}/2\pi = 1.5$ MHz, $G_{2}/2\pi = 3.5$ MHz and $\Delta_{B}=0$. For $G_{a}/2\pi$ ranging from $1$ to $4$ MHz, the most notable changes occur in the central transparency windows, which undergo substantial broadening and deepening, while the remaining windows show only minor variations. This distinct behavior reveals that the central windows originate from the optomechanically induced transparency (OMIT) mechanism and are efficiently enhanced through the photon–phonon interaction.\\
	\begin{figure} [h!] 
	\begin{center}
		\includegraphics[scale=0.35]{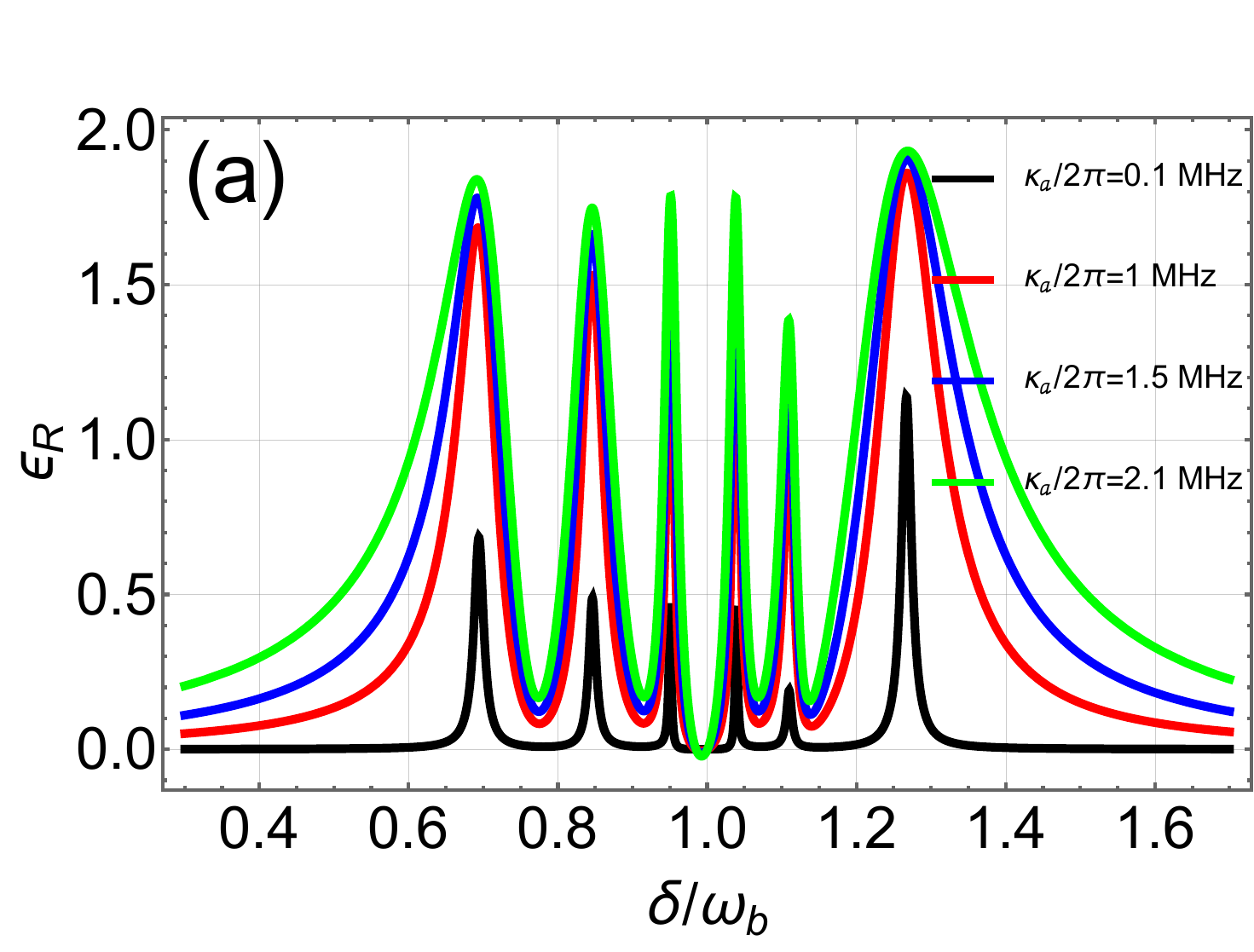}
		\includegraphics[scale=0.35]{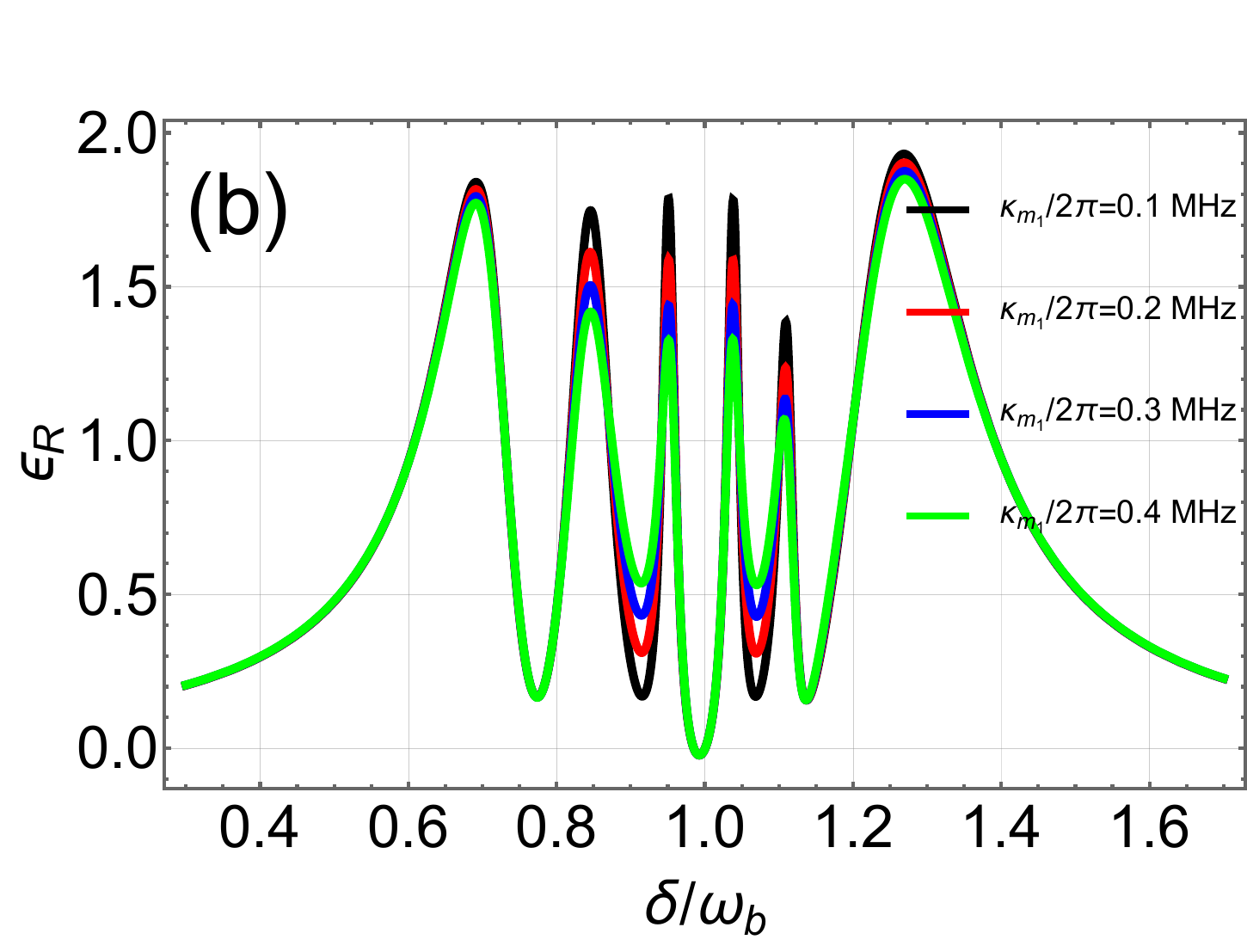}
		\includegraphics[scale=0.35]{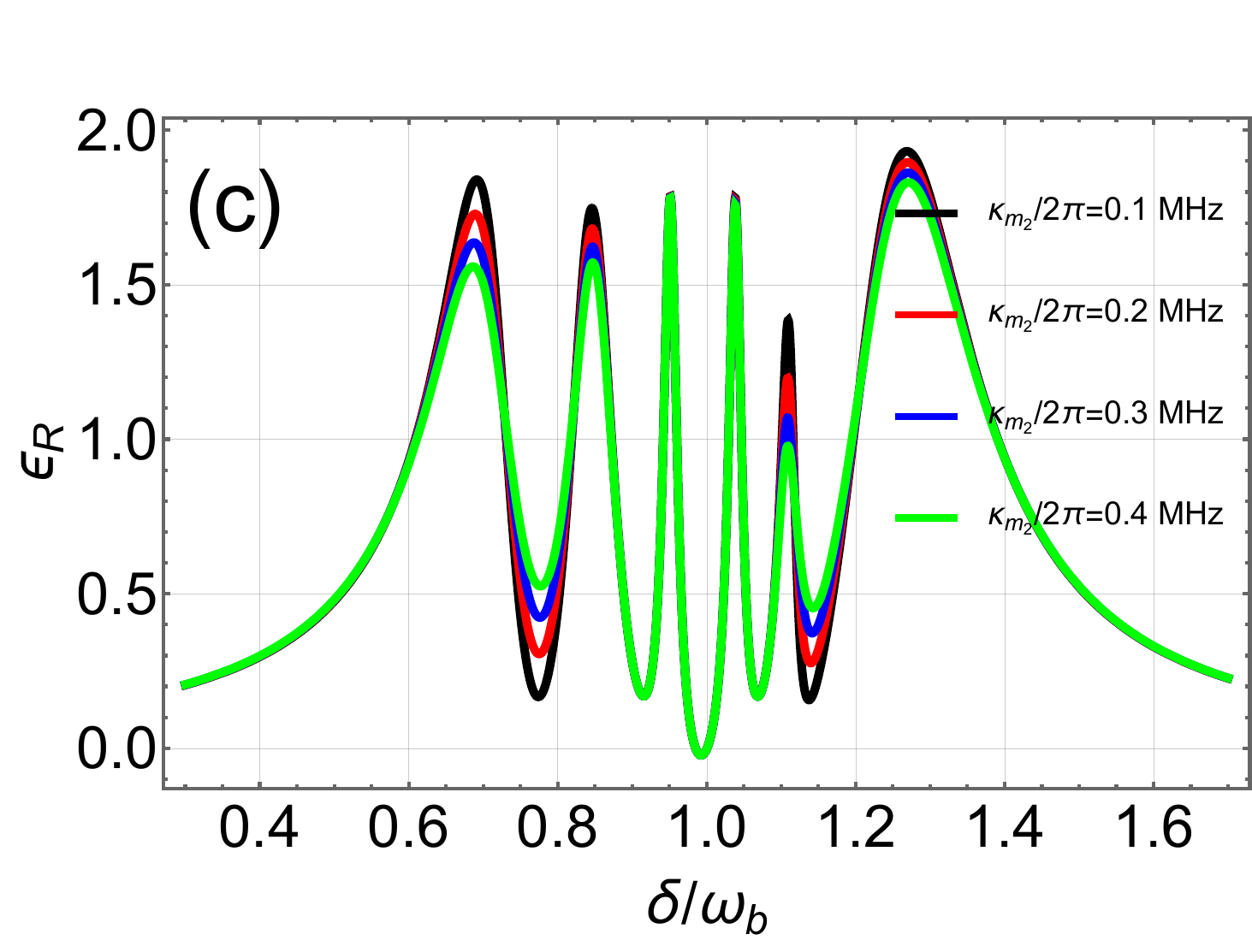}
	\caption{Real part of the output field $\epsilon_{\mathrm{R}}$ as a function of $\delta/\omega_{b}$ for different values of: (a) the cavity decay rate, (b) the dissipation rate of magnon $m_{1}$, and (c) the dissipation rate of magnon $m_{2}$, with $g_{1}/2\pi = g_{2}/2\pi=G_{1}/2\pi = 1.5~\text{MHz}$, $G_{2}/2\pi = 3.5~\text{MHz}$, and $G_{a}/2\pi = 3~\text{MHz}$. All remaining parameters are given in the Table~\ref{Tab}.}\label{l}
	\end{center}
\end{figure} 	
In Fig.~\ref{l}(a), we display the absorption spectrum $\epsilon_{\mathrm{R}}$ of the output field versus the normalized probe detuning $\delta/\omega_{b}$ for several values of the cavity decay rate.  Other parameter values are the given in the main text. From Fig.~\ref{l}(a), we clearly see that as the cavity decay rate decreases, the transparency windows become wider and deeper. This behavior is expected because a lower decay rate corresponds to a longer cavity photon lifetime, which increases the coherence of the cavity field. A more coherent cavity field interacts more effectively with the magnon modes, enhancing the quantum interference responsible for the transparency effect and resulting in broader and more pronounced transparency windows. In Figs.~\ref{l}(b) and \ref{l}(c), we plot the absorption spectrum $\epsilon_{\mathrm{R}}$ of the output field versus the normalized probe detuning $\delta/\omega_{b}$ for several values of the magnon dissipation rates. From Fig.~\ref{l}(b), we observe that the second and fourth transparency dips become deeper and wider as the dissipation rate of the first magnon $m_1$ decreases, since these transparency windows are associated with the interaction involving $m_1$. Similarly, in Fig.~\ref{l}(c), the first and fifth transparency dips become deeper and wider when the dissipation rate of the second magnon $m_2$ decreases, because these windows correspond to the interaction involving $m_2$. These results indicate that a lower magnon dissipation rate enhances the photon–magnon coupling, leading to stronger quantum interference and, consequently, more pronounced transparency windows.
	\section{SLOW AND FAST LIGHT}\label{02}
	This section focuses on controlling the transmission of the probe optical field to explore the appearance of slow and fast light effects in a magnomechanical system. Starting from Eq.~\eqref{s1}, the normalized transmitted probe field can be written as
	\begin{equation}\label{s1}
		\mathcal{T} = \frac{\varepsilon_p - 2\kappa_a a_-}{\varepsilon_p}.
	\end{equation}
	The phase $\phi$ of the transmitted probe signal is defined as
	\begin{equation}\label{s2}
		\phi = \mathrm{Arg}(\mathcal{T}).
	\end{equation}
	This phase, introduced in Eq.~\eqref{s2}, is a key quantity for evaluating the group delay, which is defined as 
	\begin{equation}\label{cv}
		\tau = \frac{\partial \phi}{\partial \omega_p}
		= \mathrm{Im}\!\left[ \mathcal{T}^{-1} \frac{\partial \mathcal{T}}{\partial \omega_p} \right],
	\end{equation}
	where $\tau$ represents the temporal delay experienced by the output field. A positive value of the group delay ($\tau > 0$) indicates the presence of slow light propagation, while a negative value ($\tau < 0$) signifies the occurrence of fast light behavior.\\
	
\begin{figure} [h!] 
		\begin{center}
			\includegraphics[scale=0.4]{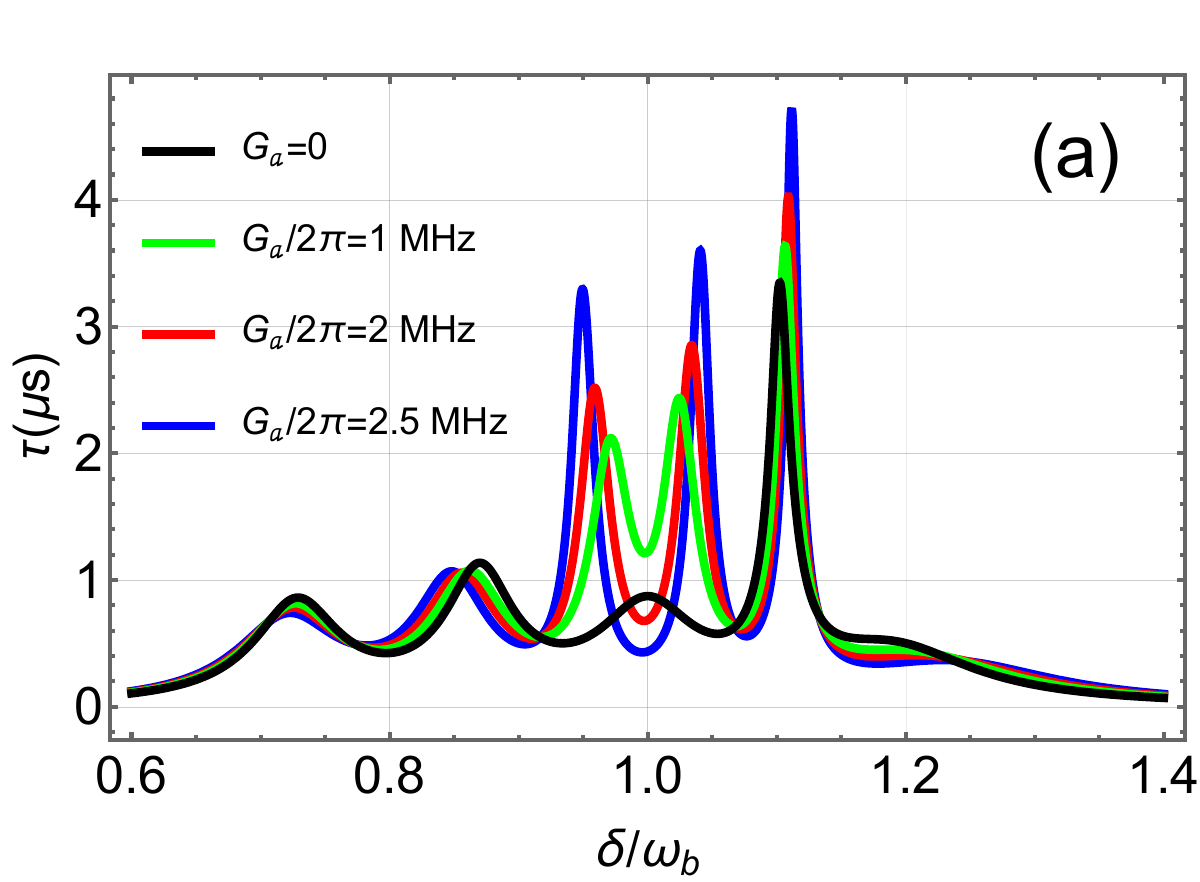}
			\includegraphics[scale=0.42]{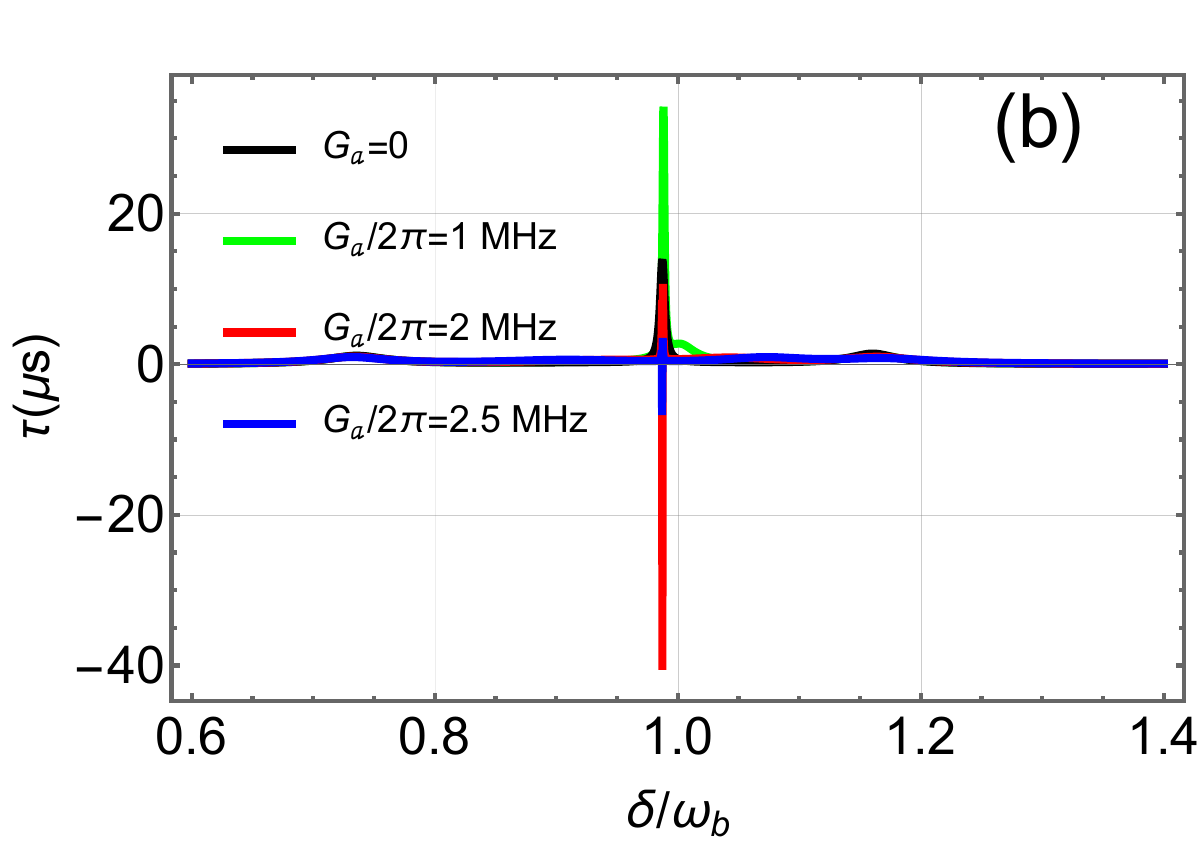}
			\includegraphics[scale=0.4]{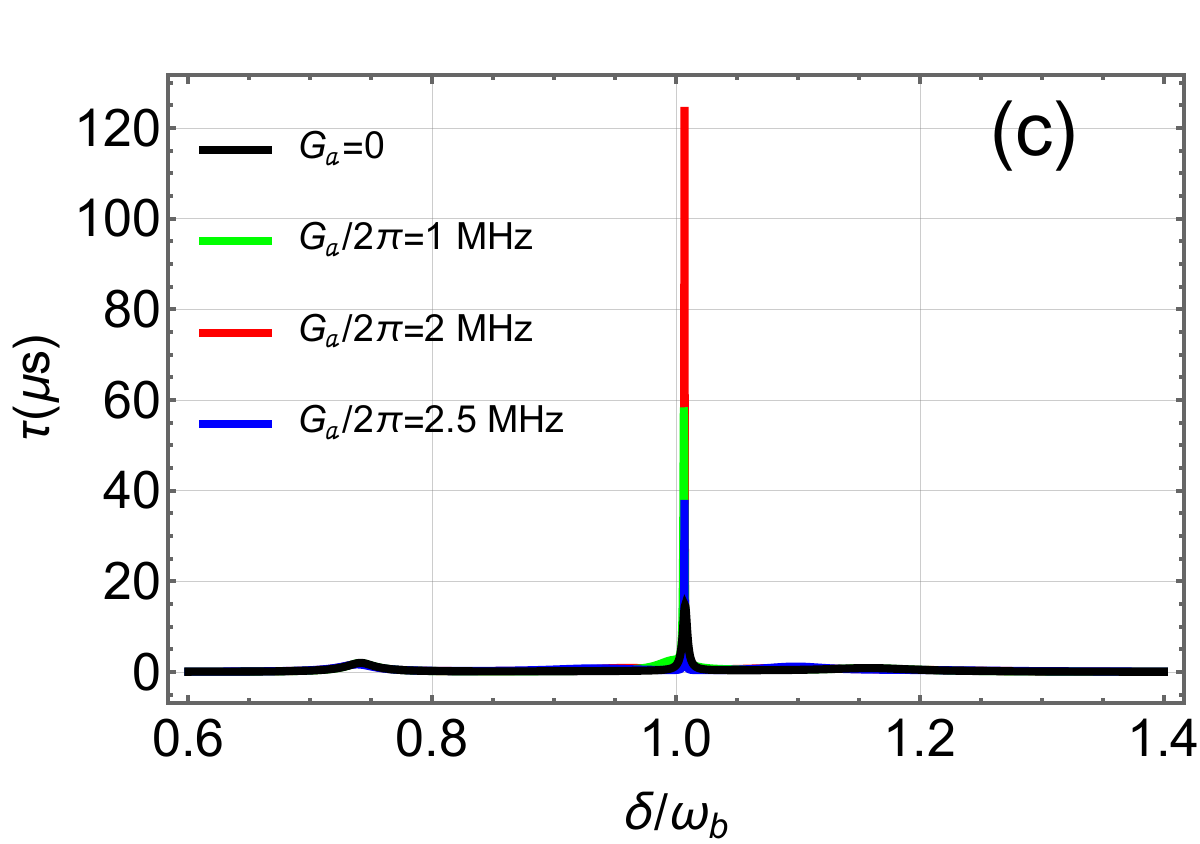}
			\caption{Group delay $\tau$ as a function of the normalized detuning $\delta/\omega_b$ for several values of the photon-phonon coupling $G_a$. Panels (a) $\Delta_{B}=0$, (b) $\Delta_{B}=-0.5\omega_{b}$, and (c) $\Delta_{B}=0.5\omega_{b}$, respectively. The remaining parameters are given in Section \ref{0}.}\label{w1}
		\end{center}
	\end{figure}

	Fig.~\ref{w1} illustrates the group delay $\tau$ of the output field at the probe frequency as a function of the normalized detuning $\delta/\omega_b$, for different values of the photon–phonon coupling strength $G_a$, while all other parameters are kept fixed. In Fig.~\ref{w1}(a), where the Barnett effect is deactivated ($\Delta_B = 0$), the group delay remains strictly positive over the considered detuning range, indicating a clear slow light regime. Moreover, the slow light effect increases with increasing photon–phonon coupling $G_a$, demonstrating that stronger optomechanical interaction of the mirror enhances the group delay. Figure~\ref{w1}(b) corresponds to a negative Barnett effect, $\Delta_B = -0.5\omega_b$. In this case, the group delay at $\delta = \omega_b$ is $\tau = 16.3~\mu\text{s}$ for $G_a = 0$, confirming the presence of slow light even without photon–phonon coupling. When the coupling is introduced, the slow light effect is significantly enhanced: for $G_a/2\pi = 1$ MHz and $2$ MHz, the group delay increases to $\tau = 61.5~\mu\text{s}$ and $127.5~\mu\text{s}$, respectively. However, for a stronger coupling of $G_a/2\pi = 2.5$ MHz, the group delay decreases to $\tau = 40.9~\mu\text{s}$. Figure~\ref{w1}(c) shows the case of a positive Barnett effect, $\Delta_B = 0.5,\omega_b$. For $G_a = 0$ and $G_a/2\pi = 1$ MHz, the group delay remains positive, reaching $\tau = 15.2~\mu\text{s}$ and $\tau = 35.92~\mu\text{s}$, respectively. When the coupling increases to $G_a/2\pi = 2$ MHz, the positive group delay is reduced to $\tau = 12.5~\mu\text{s}$. Notably, a negative group delay of $\tau = -40.4~\mu\text{s}$ appears at $\delta = 0.98,\omega_b$, revealing the emergence of a fast light regime. For a further increase to $G_a/2\pi = 2.5$ MHz, the positive delay decreases to $\tau = 5.1~\mu\text{s}$, while the negative delay is reduced to $\tau = -9~\mu\text{s}$. By comparing Figs.~\ref{w1}(b) and~\ref{w1}(c), we observe that when the Barnett effect changes from negative to positive, the positive group delay decreases significantly for the same values of the photon–phonon coupling $G_a$. For example, for $G_a/2\pi = 1$ MHz and $\Delta_B = -0.5\omega_b$, the group delay is $\tau = 61.5~\mu\text{s}$, whereas it decreases to $\tau = 35.92~\mu\text{s}$ for $G_a/2\pi = 1$ MHz and $\Delta_B = 0.5\omega_b$. Moreover, for stronger coupling ($G_a/2\pi = 2$ MHz), the negative Barnett effect leads to a large positive delay ($\tau = 127.5~\mu\text{s}$), while the positive Barnett effect results in a negative delay ($\tau = -40.4~\mu\text{s}$), indicating the emergence of the fast light regime. These results indicate that the group delay $\tau$ can be efficiently tuned by varying both the photon-phonon coupling $G_{a}$ and the Barnett effect $\Delta_B$, allowing precise control over the transition between slow and fast light regimes.\\
			\begin{figure} [h!] 
		\begin{center}
			\includegraphics[scale=0.5]{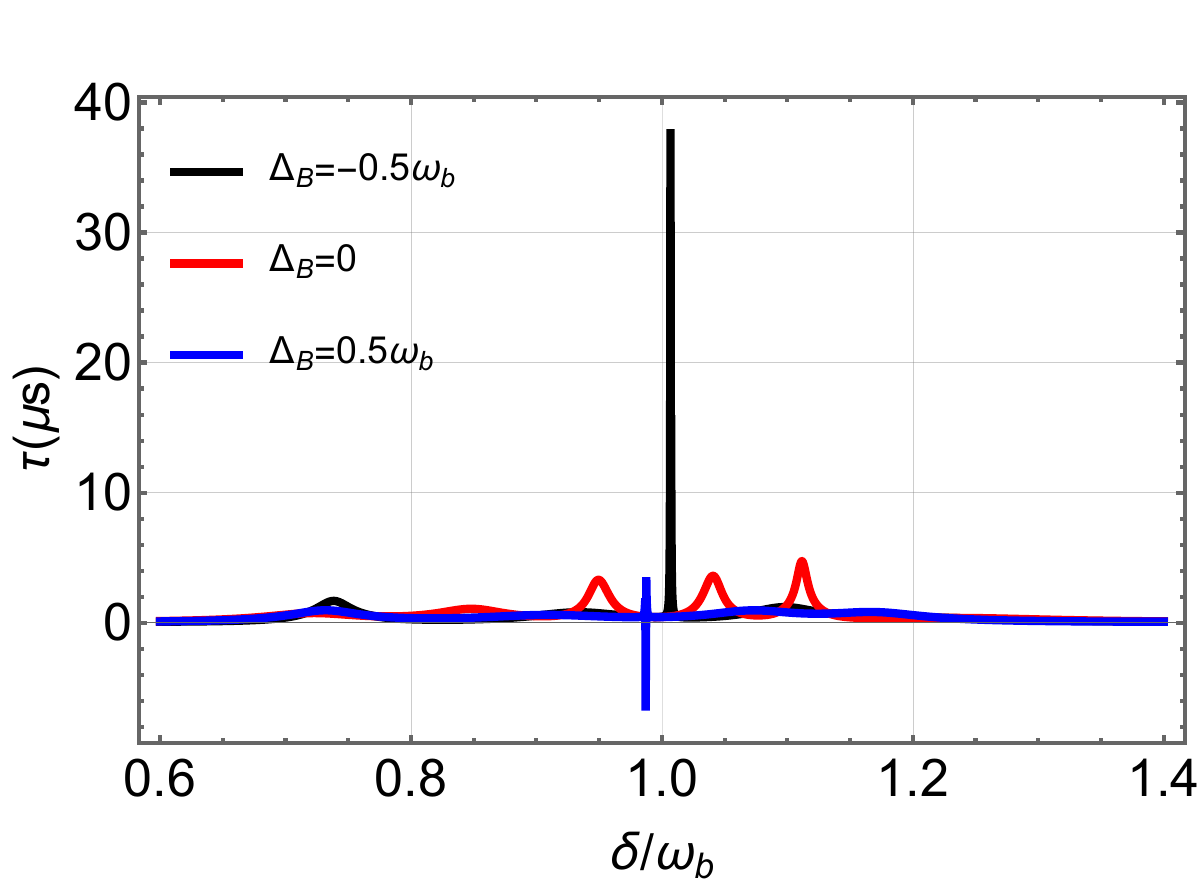}
			\caption{Group delay $\tau$ as a function of the normalized detuning $\delta/\omega_b$ for several values of the Barnett effect $\Delta_{B}$. The remaining parameters are given in Section \ref{0}.}\label{w2}
		\end{center}
	\end{figure} 
	Fig.~\ref{w2} illustrates the variation of the group delay $\tau$ as a function of the normalized detuning $\delta/\omega_b$ for different values of the Barnett effect $\Delta_B$. In the absence of the Barnett effect ($\Delta_B = 0$), the system exhibits symmetric resonant features around $\delta/\omega_b \approx 1$, leading to moderate positive group delay peaks that indicate the emergence of slow-light behavior. When a negative Barnett shift is introduced ($\Delta_B = -0.5\omega_b$), a sharp and pronounced enhancement of the group delay is observed near resonance, reaching significantly larger values, which reflects a strong amplification of the slow-light effect. In contrast, for a positive Barnett effect ($\Delta_B = 0.5\omega_b$), the group delay is reduced and even becomes negative near the resonance point, indicating a transition toward fast-light propagation. These results demonstrate that both the magnitude and the sign of the group delay can be effectively controlled by tuning the Barnett effect, providing a flexible mechanism for switching between slow- and fast light regimes.
    \section{Nonreciprocal}\label{03}
    In this section, we investigate the nonreciprocal behavior of the absorption and the group delay induced by the Barnett effect, which plays a crucial role in controlling directional signal propagation and realizing integrated nonreciprocal devices. Since the expressions for the absorption spectrum $\epsilon_R$ and the group delay $\tau$ have already been derived in Eqs.~\eqref{cp} and~\eqref{cv}, respectively, we directly analyze their dependence on the sign of the Barnett shift $\Delta_B$.
    \subsection{Nonreciprocal absorption}
\begin{figure} [h!] 
		\begin{center}
			\includegraphics[scale=0.6]{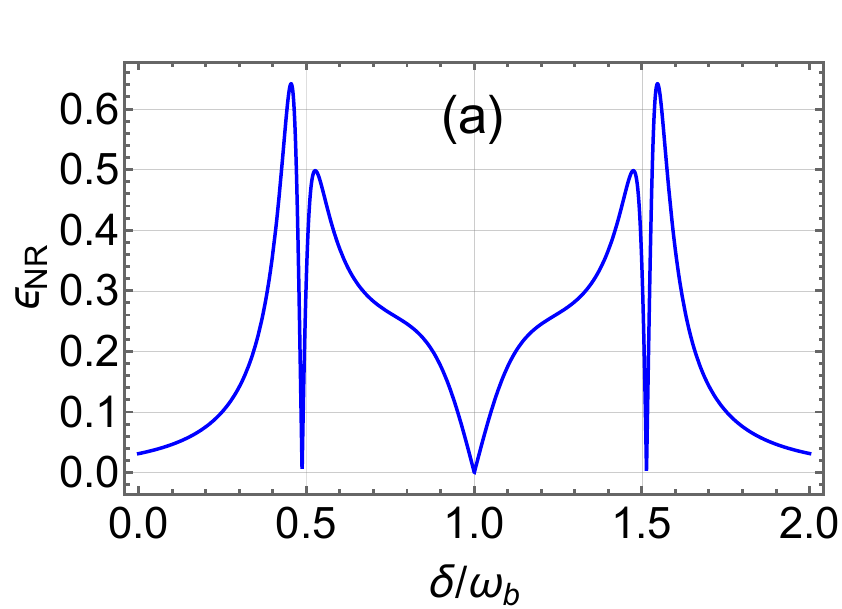}
			\includegraphics[scale=0.525]{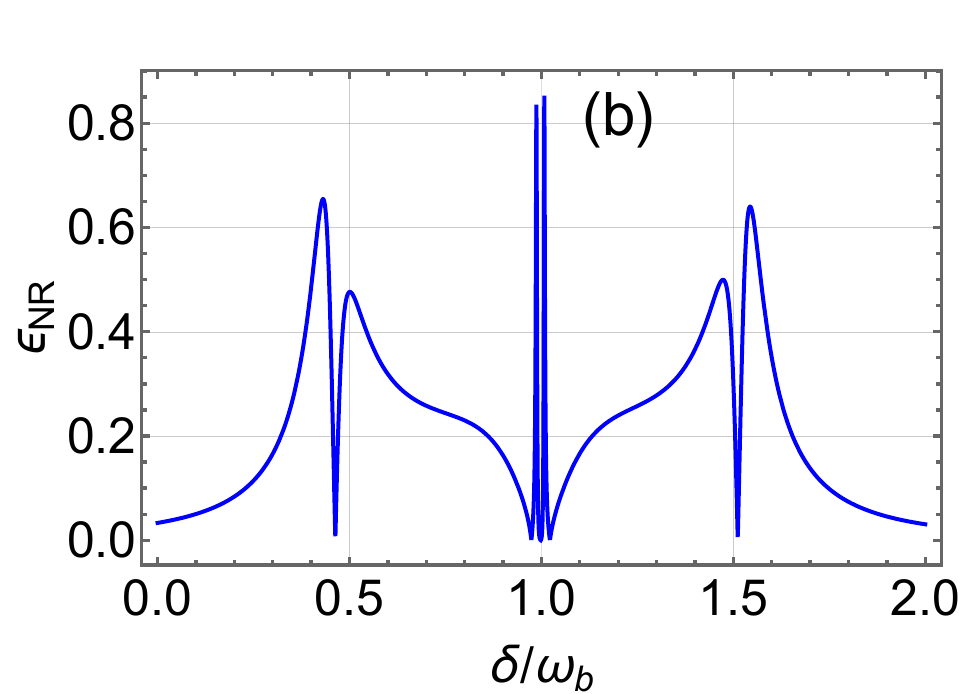}
			\includegraphics[scale=0.528]{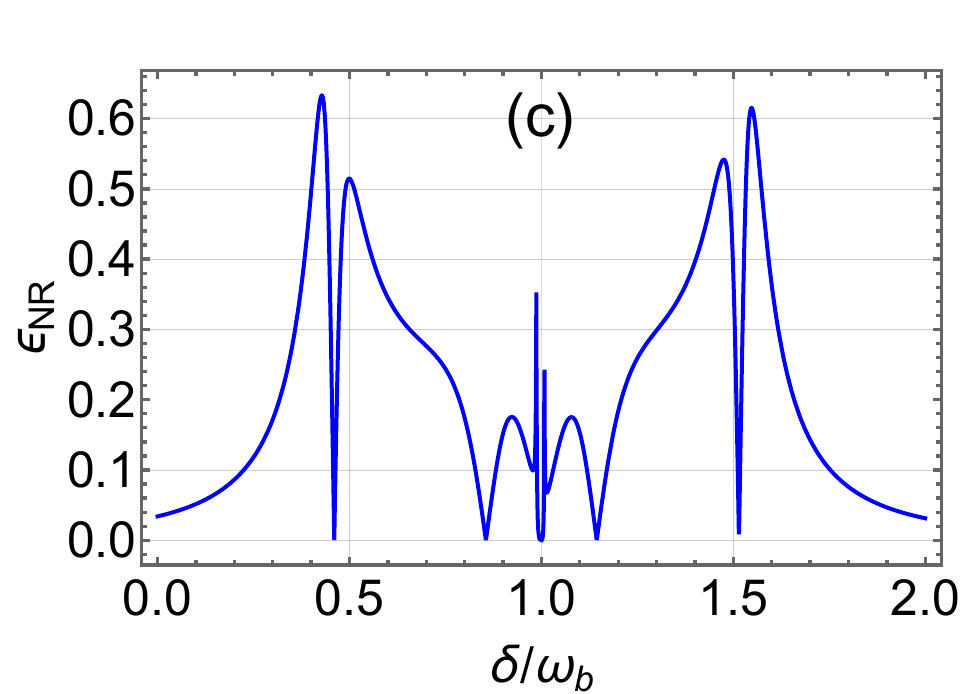}
			\caption{ The nonreciprocal absorption as a function of $\delta/\omega_{b}$. (a) $g_{2}=G_{1}=G_{2}=G_a=0$, $g_{1}/2\pi = 1.5$ MHz and $\Delta_{B}=0.5\omega_{b}$; (b) $g_{2}=G_{2}=G_a=0$ and $g_{1}/2\pi = G_{1}/2\pi = 1.5$ MHz;	(c) $G_{2}=G_a=0$ with $g_{1}/2\pi = g_{2}/2\pi = G_{1}/2\pi = 1.5$ MHz and $|\Delta_{B}|=0.5\omega_{b}$. All remaining parameters are given in Section \ref{0}.}\label{s}
		\end{center}
	\end{figure}

    To quantify the Barnett effect induced nonreciprocal absorption, we introduce the contrast ratio $\epsilon_{NR}$ (with $0 \leq \epsilon_{NR} \leq 1$), defined as
   \begin{equation}
	\epsilon_{NR}=\frac{\left|\epsilon_{R}\left(\Delta_B<0\right)-\epsilon_{R}\left(\Delta_B>0\right)\right|}{\epsilon_{R}\left(\Delta_B<0\right)+\epsilon_{R}\left(\Delta_B>0\right)}.
   \end{equation}
    Here, $\epsilon_R$ represents the real part of the output field response, which characterizes the absorption spectrum. A contrast ratio $\epsilon_{NR}=0$ indicates reciprocal absorption, while $\epsilon_{NR}=1$ corresponds to ideal nonreciprocal absorption. Fig.~\ref{s} clearly shows that the nonreciprocal absorption can be actively controlled and even switched on or off by tuning the normalized detuning $\delta/\omega_b$. In particular, the contrast ratio $\epsilon_{NR}$ can be continuously tuned between $0$ and $1$ by varying $\delta/\omega_b$, enabling the realization of optimal nonreciprocal absorption together with tunable transparency windows. In Fig.~\ref{s}(a), the optimal nonreciprocity is achieved in the ranges $0 < \delta/\omega_b < 0.5$ and $1.5 < \delta/\omega_b < 2$. For Fig.~\ref{s}(b), it occurs within the narrow intervals $0.97 \le \delta/\omega_b \le 1$ and $1 \le \delta/\omega_b \le 1.03$. In Fig.~\ref{s}(c), the maximum nonreciprocity is observed approximately in the regions $0 < \delta/\omega_b < 0.46$ and $1.51 < \delta/\omega_b < 2$. By comparing Figs.~\ref{b}(b), \ref{b}(c), and \ref{b}(d) with Figs.~\ref{s}(a), \ref{s}(b), and \ref{s}(c), respectively, we observe that when the blue line ($\Delta_{B}>0$) and the black line ($\Delta_{B}<0$) overlap, no nonreciprocal effect occurs, whereas when the two lines are well separated, perfect nonreciprocity is achieved. These results indicate that, by carefully tuning the detuning $\delta$, the transparency windows and absorption spectra for opposite rotation directions can be made increasingly distinct and asymmetric, demonstrating a highly controllable and tunable nonreciprocal behavior in the system.
    
    \subsection{Nonreciprocal group delay}
\begin{figure} [h!] 
	\begin{center}
		\includegraphics[scale=0.5]{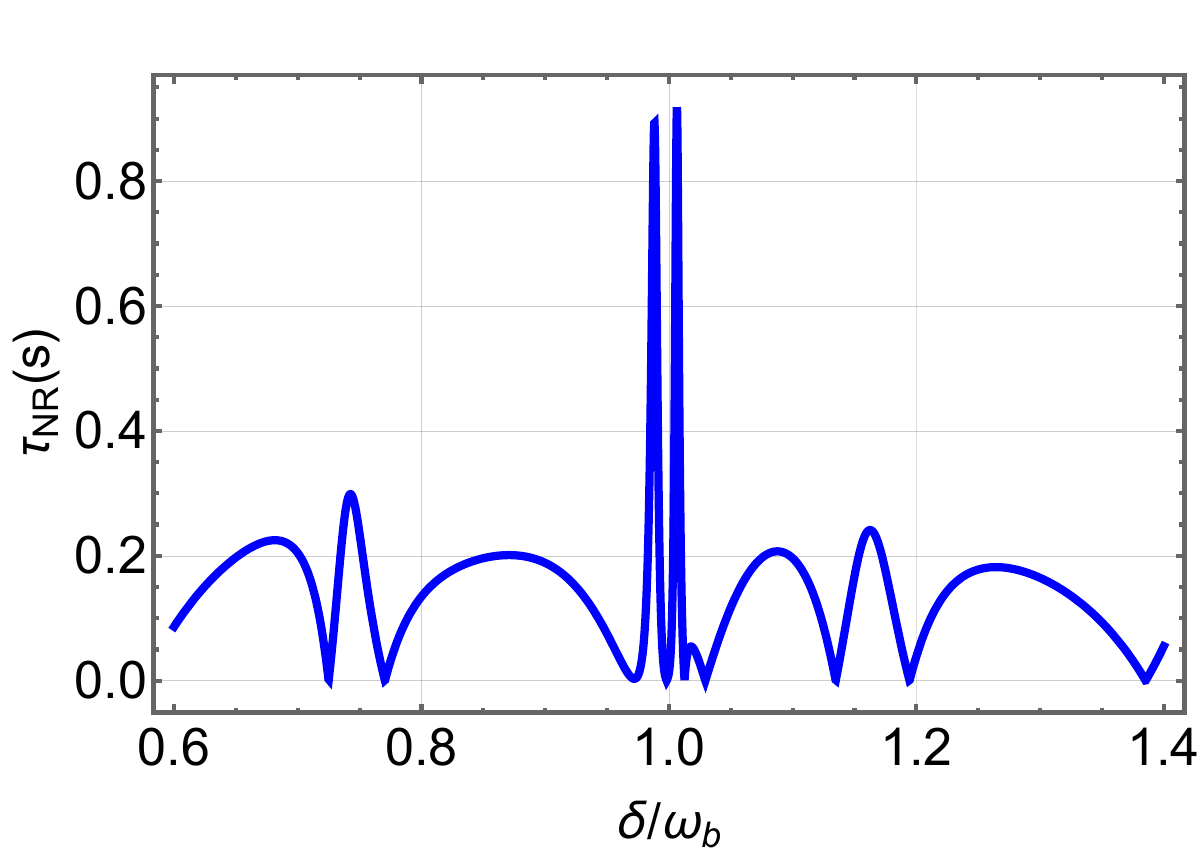}
		\caption{ The nonreciprocal group delay as a function of $\delta/\omega_{b}$, with $ G_{a}/2\pi = 1$ MHz and $|\Delta_{B}|=0.5\omega_{b}$. All remaining parameters are given in Section \ref{0}.}\label{s8}
	\end{center}
	\end{figure}

   We now turn to the nonreciprocal behavior of the group delay, whose analytical expression is provided in Eq.~\eqref{cv}. Because the phase dispersion of the output field depends sensitively on the hybrid photon–magnon–phonon interactions, reversing the sign of $\Delta_B$ changes the slope of the phase response and thus the group delay. To characterize this directional asymmetry, we define the group delay nonreciprocity factor as
     \begin{equation}
	\tau_{NR}=\frac{\left|\tau\left(\Delta_B<0\right)-\tau\left(\Delta_B>0\right)\right|}{\tau\left(\Delta_B<0\right)+\tau\left(\Delta_B>0\right)},
    \end{equation}
    where $\tau_{NR}=0$ indicates reciprocal group delay, while $\tau_{NR}=1$ corresponds to ideal nonreciprocal group delay. Fig.~\ref{s8} demonstrates that the nonreciprocity of the group delay can be actively switched on or off by tuning the auxiliary cavity detuning $\delta/\omega_{b}$. The bidirectional contrast ratio for the group delay can be continuously varied between $0$ and $1$ by adjusting $\Delta_c$. Specifically, nonreciprocal group delay is observed within the ranges $0.97 \le \delta/\omega_b \le 0.99$ and $1 \le \delta/\omega_b \le 1.01$. Therefore, the interplay between the Barnett effect and hybrid magnon–phonon–photon interactions enables simultaneous control of nonreciprocal absorption and nonreciprocal slow/fast light, offering a flexible platform for directional photonic devices and on-chip signal processing.
	\section{FEASIBILITY}\label{04}
   The experimental feasibility of the proposed model is supported by recent advances in cavity magnomechanical and optomechanical systems. Experimental studies have confirmed that magnomechanically induced transparency arises from quantum side band interference due to coherent coupling between magnon and phonon modes \cite{0}. Collective magnon excitations in yttrium iron garnet (YIG) crystals embedded in microwave cavities interact strongly with microwave photons via magnetic dipole coupling \cite{25}, while magnetostrictive effects provide an efficient coupling channel between magnons and phonons \cite{0}. Moreover, the optical cavity required for optomechanical interactions can be realized by attaching a small, highly reflective mirror to a micro-bridge structure \cite{K2}. Radiation pressure induced optomechanical coupling has already been experimentally demonstrated, leading to optomechanically induced transparency (OMIT) \cite{K3}. Based on these established techniques, coherent state-swap interactions among magnons, phonons, and photons can be effectively implemented \cite{13}. Taken together, these experimental achievements indicate that all essential ingredients of our proposal have been realized individually, confirming that the present scheme is feasible with current experimental technology.
	
    \section{CONCLUSION}\label{05}
    In this work, we have theoretically explored magnomechanically induced transparency (MMIT), Fano resonances, slow and fast light effects, as well as nonreciprocal absorption and group delay in a cavity magnomechanical system. By examining the response of a weak probe field under the action of a strong control field, we demonstrated that the absorption spectrum exhibits five transparency windows. These windows arise from the hybridization of photon-phonon, photon-magnon, and phonon-magnon interactions. We further analyzed the impact of the Barnett effect on both the absorption and dispersion properties of the system. Our findings reveal that appropriate tuning of the Barnett effect significantly modifies the transparency structure and induces controllable Fano resonances, offering additional flexibility in engineering the spectral response. The roles of the cavity decay rate and the magnon dissipation rate were also clarified, showing how these parameters influence the line widths and contrast of the transparency windows. Moreover, we showed that the propagation of light can be efficiently manipulated by controlling the photon-phonon coupling strength of the membrane together with the Barnett effect. Finally, we investigated nonreciprocal absorption and group delay under different coupling regimes. The results confirm that strong nonreciprocal behavior can be achieved through suitable parameter adjustment, leading to asymmetric absorption and directional dependent group delay.
\section*{Acknowledgments}

M.~Amghar gratefully acknowledges financial support provided by the National Center for Scientific and Technical Research (CNRST) under the PhD-Associate Scholarship (PASS) program.
	\appendix
	\renewcommand{\thesection}{ \Alph{section}}
	\section{Derivation of $a_-$} \label{AP}
	$$h_1=\kappa_a+i(\tilde{\Delta}_a-\delta),\quad h_2=\kappa_{m_1}+i(\tilde{\Delta}_{m_1}-\delta),\quad h_3=\omega_{b_1}-\frac{\delta}{\omega_{b_1}}(\delta+i\gamma_1),\quad h_4=\kappa_{m_1}-i(\tilde{\Delta}_{m_1}+\delta),\quad h_5=\kappa_a-i(\tilde{\Delta}_a+\delta),$$
	$$h_6=\kappa_{m_2}-i(\tilde{\Delta}_{m_2}+\delta),\quad h_7=\omega_{b_2}-\frac{\delta}{\omega_{b_2}}(\delta+i\gamma_2),\quad h_8=\kappa_{m_2}+i(\tilde{\Delta}_{m_2}-\delta),\quad h_9=\omega_{b_3}-\frac{\delta}{\omega_{b_3}}(\delta+i\gamma_3)$$
	$$A=1+\frac{G_{22}^2}{ih_7h_8},\quad B=1-\frac{G_{22}^2}{ih_6h_7A},\quad E=1+\frac{g_2^2}{h_5h_6B}-\frac{G_{aa}^2}{ih_5h_9},\quad F=\frac{ig_2^2G_{22}^2}{h_5h_6h_7h_8AB}-\frac{G_{aa}^2}{ih_5h_9},\quad G=1+\frac{g_1^2}{h_4h_5E} $$
	$$K=1-\frac{G_{11}^2}{ih_3h_4G},\quad L=1+\frac{G_{11}^2}{ih_2h_3K},\quad M=\frac{-ig_1}{h_2}+\frac{g_1G_{11}^2F}{h_2h_3h_4EGK},\quad X_1=1+\frac{G_{11}^2}{ih_2h_3},\quad X_2=1-\frac{G_{11}^2}{ih_3h_4X_1};\quad $$
	$$X_3=1+\frac{g_1^2}{h_4h_5X_2}-\frac{G_{aa}^2}{ih_5h_9},\quad X_4=\frac{ig_1^2G_{11}^2}{h_2h_3h_4h_5X_1X_2}-\frac{G_{aa}^2}{ih_5h_9},\quad X_5=1+\frac{g_2^2}{h_5h_6X_3},\quad X_6=1-\frac{G_{22}^2}{ih_6h_7X_5},\quad X_7=1+\frac{G_{22}^2}{ih_7h_8X_6},$$
	$$X_8=\frac{-ig_2}{h_8}+\frac{g_2G_{22}^2X_4}{h_6h_7h_8X_3X_5X_6},\quad Z_1=1+\frac{g_1^2}{h_4h_5X_2}+\frac{g_2^2}{h_5h_6B},\quad Z_2=\frac{ig_1^2G_{11}^2}{h_2h_3h_4h_5X_1X_2}+\frac{ig_2^2G_{22}^2}{h_5h_6h_7h_8AB},\quad Z_3=1-\frac{G_{aa}^2}{ih_5h_9Z_1}$$
	$$Z_4=\frac{-G_{aa}}{ih_9}+\frac{G_{aa}Z_2}{ih_9Z_1}.$$
	Here, $G_{11}=G_1/\sqrt{2}$, $G_{22}=G_2/\sqrt{2}$ and $G_{aa}=G_a/\sqrt{2}$, with $G_r=i\sqrt{2}G_{0r}m_{rs}$ ($r=1,2$) and $G_a=i\sqrt{2}g_aa_{s}$, representing the effective opto-magnomechanical coupling rate, where $|\tilde{\Delta}_{m_1}|,|\tilde{\Delta}_{m_2}|,\left|\Delta_a\right| \gg \kappa_a, \kappa_{m_1},\kappa_{m_2}$.\\
	\section{Rabi frequency of the magnon mode $m_1$} \label{A}
	The Rabi frequency $\Omega_l$ characterizes the coupling strength between the driving magnetic field and the magnon mode. We now derive its explicit form, which is essential for determining the effective magnomechanical coupling rate and validating the linearized model. The Hamiltonian for a single spin $\vec{s}$ in an external magnetic field $\vec{B}$ is  
	\begin{equation}
		H = - \gamma \vec{s} \cdot \vec{B},
	\end{equation}
	where $\gamma$ is the gyromagnetic ratio. For a YIG sphere containing a large number of spins, it is convenient to consider the total spin contribution by introducing the collective spin angular momentum  
	\begin{equation}
		\vec{S} = \sum \vec{s}.
	\end{equation}
	
	When this collective spin interacts with an oscillating magnetic field along the $y$-axis, with amplitude $B$ and frequency $\omega_{L}$, the interaction Hamiltonian reads  
	\begin{equation}
		H_{L} = - \gamma \vec{S} \cdot \vec{B} = - \gamma S_y B \cos(\omega_{L} t),
	\end{equation}
	where $\vec{S} = (S_x, S_y, S_z)$. Using the spin raising and lowering operators, $S^\pm = S_x \pm i S_y$, we can rewrite this as  
	\begin{equation}
		H_{L} = i \frac{\gamma B}{4} (S^+ - S^-) \left( e^{i \omega_{L} t} + e^{-i \omega_{L} t} \right).
	\end{equation}
	
	Applying the Holstein-Primakoff transformation, the collective spin operators $S^\pm$ are related to the magnon annihilation and creation operators $m$ and $m^\dag$ by  
	\begin{equation}
		S^+ = \hbar \, m_1 \sqrt{2Ns - m_1^\dag m_1}, \quad 
		S^- = \hbar \, m_1^\dag \sqrt{2Ns - m_1^\dag m_1},
	\end{equation}
	where $s = 5/2$ is the spin of a ground-state Fe$^{3+}$ ion in YIG and $N$ is the total number of spins. In the low-excitation limit, $\langle m_1^\dag m_1 \rangle \ll 2Ns$, these approximate to  
	\begin{equation}
		S^+ \approx \hbar \,m_1 \sqrt{5N} , \quad
		S^- \approx \hbar \, m_1^\dag\sqrt{5N} .
	\end{equation}
	
	Substituting into $H_{L}$ yields  
	\begin{equation}
		\begin{split}
			H_{L} &= i \hbar \frac{\sqrt{5}}{4} \gamma \sqrt{N} B_l \, (m_1 - m_1^\dag) \left( e^{i \omega_{L} t} + e^{-i \omega_{L} t} \right) \\
			&\approx i \hbar \Omega \left( m_1 e^{i \omega_{L} t} - m_1^\dag e^{-i \omega_{L} t} \right),
		\end{split}
	\end{equation}
	where the Rabi frequency is  
	\begin{equation}
		\Omega= \frac{\sqrt{5}}{4} \gamma \sqrt{N} B, \quad
		\gamma / 2\pi = 28~\text{GHz/T}, \quad
		N = \nu \mathcal{V}.
	\end{equation}
	Here, $\mathcal{V}$ is the volume of the YIG sphere and $\nu = 4.22 \times 10^{27}~\text{m}^{-3}$ is the spin density of YIG.
	\section{Negligibility of Kerr nonlinearities in the YIG sphere}
	
	In the analysis above, the Kerr nonlinear term $K m^\dag m m^\dag m$ in the Hamiltonian can be safely neglected \cite{47,K1}. Here, $K$ denotes the Kerr coefficient, which scales inversely with the system volume and characterizes intrinsic material nonlinearities. This term may arise due to the strong magnon drive, but its effect is extremely small. For a YIG sphere with a diameter of $1$~mm, the Kerr coefficient has been estimated as $K \sim 10^{-10} / 2\pi~\text{Hz}$ \cite{47,K1}. Reducing the diameter to $250~\mu\text{m}$ increases $K$ roughly sixfold, giving $K \sim 6.4 \times 10^{-9} /2\pi~\text{Hz}$. Considering the parameters used in the figures, the nonlinear contribution $K |\langle m \rangle|^3$ is on the order of $5.7 \times 10^{13}~\text{Hz}$, which remains much smaller than $\Gamma_l \ge 2.23 \times 10^{14}~\text{Hz}$, with $l = 1,2$. The large difference between these scales confirms that Kerr nonlinearities can be safely neglected in our analysis.

\end{document}